\documentclass[twocolumn,aps,showpacs,superscriptaddress,longbibliography]{revtex4-1}

\usepackage{graphicx}
\usepackage{color}
\usepackage{latexsym}

\begin{document}

\title{Two-beam light with simultaneous anticorrelations in photon-number fluctuations and sub-Poissonian statistics}

\author{Jan Pe\v{r}ina Jr.}
\email{jan.perina.jr@upol.cz} \affiliation{Joint Laboratory of Optics of
Palack\'{y} University and Institute of Physics of the Czech Academy of
Sciences, Faculty of Science, Palack\'{y} University, 17. listopadu 12, 77146
Olomouc, Czech Republic}

\author{V\'{a}clav Mich\'{a}lek}
\affiliation{Joint Laboratory of Optics of Palack\'{y} University and Institute
of Physics of the Czech Academy of Sciences, Faculty of Science, Palack\'{y}
University, 17. listopadu 12, 77146 Olomouc, Czech Republic}

\author{Radek Machulka}
\affiliation{Institute of Physics of the Czech Academy of Sciences, Joint
Laboratory of Optics of Palack\'{y} University and Institute of Physics of CAS,
17. listopadu 50a, 772 07 Olomouc, Czech Republic}

\author{Ond\v{r}ej Haderka}
\affiliation{Institute of Physics of the Czech Academy of Sciences, Joint
Laboratory of Optics of Palack\'{y} University and Institute of Physics of CAS,
17. listopadu 50a, 772 07 Olomouc, Czech Republic}

\begin{abstract}
Two twin beams with a shared signal beam and separated idler beams are used
together with the photon-number-resolving postselection in the signal beam to
arrive at two coupled beams with anticorrelations in photon-number
fluctuations. Moreover, the beams exhibit the sub-Poissonian photon-number
statistics in their marginal distributions under suitable conditions. The
postselected fields with the increasing mean photon numbers are reconstructed
from the experimental photocount histograms by the maximum likelihood approach.
Also a suitable Gaussian fit of both original twin beams and simulation of the
postselection process are applied to arrive at the corresponding photon-number
distributions. Their nonclassical properties are analyzed by suitable
nonclassicality criteria and quantified by the corresponding nonclassicality
depths. Determining the appropriate quasi-distributions of integrated
intensities with negative values, the performance of different nonclassicality
criteria is judged. Properties of the postselected fields reached both by the
used and ideal photon-number-resolved detectors are mutually compared.
\end{abstract}

\maketitle

\section{Introduction}

Twin beams (TWBs) generated in spontaneous parametric down-conversion
\cite{Boyd2003} are endowed with highly nonclassical properties
\cite{Mandel1995} observed in different degrees of freedom. Their entanglement
occurring in the polarization degrees of freedom has been exploited to test the
quantum mechanics via the violation of the Bell inequalities
\cite{Weihs1998,Genovese2005} or to teleport the polarization state of a photon
\cite{Bouwmeester1997}. Tight spatial correlations of the photons in a TWB lie
in the heart of quantum imaging \cite{Genovese2016}. On the other hand perfect
correlations in photon numbers of the signal and idler beams, that constitute a
TWB, \cite{Jedrkiewicz2004,Haderka2005a,Bondani2007,Blanchet2008,Brida2009a}
gave rise to the method of absolute detector calibration
\cite{Klyshko1980,Brida2006a,PerinaJr2012a,Haderka2014}.

Also, a very efficient method for sub-Poissonian light generation by
photon-number-resolved postselection (in cw regime:
\cite{Rarity1997,Laurat2003,Zou2006}, in pulsed regime:
\cite{Bondani2007,PerinaJr2013b,Lamperti2014,Iskhakov2016,Harder2016}) is based
upon TWBs. Such states represent a generalization of (heralded) single-photon
Fock states
\cite{Zeldovich1969,PerinaJr2001,Alibard2005,Brida2012,Horoshko2019} to more
intense fields described in the Hilbert spaces of larger dimensions. Such
fields then allow, among others, to increase the capacity of communication
channels \cite{Saleh1987}. The highly-nonclassical single-photon Fock states
are a workhorse of the broad area of quantum-information processing
\cite{Nielsen2000} based on the discrete variables. They also find their
application in sub-shot-noise imaging
\cite{Jakeman1986,Brida2010a,Whittaker2017,Li2018,SabinesChesterkind2019}.

The used postselection process represents a critical step in the preparation of
highly-nonclassical states as it degaussifies the original Gaussian TWB. This
makes the postselection method very prospective for the generation of more
complex quantum states potentially needed in future quantum-information
protocols that will go beyond the single-photon Fock states. Also the
application of such states in quantum metrology
\cite{Abouraddy2002,Brida2010a,Giovannetti2006,Giovannetti2011} is expected. We
note that the generation of photon-number-subtracted states
\cite{Agarwal1992,Iskhakov2016a,Barnett2018} represents a special variant of
the postselection with photon-number-resolving detectors that allows to
generate various kinds of nonclassical states, even from TWBs
\cite{Kim2005,MaganaLoaiza2019}.

To put our considerations about the states with different photon numbers and
photon-number correlations into the general context, we remind the reader that,
according to the second-quantization of electromagnetic fields in the quantum
mechanics \cite{Mandel1995}, any state of an optical field can be decomposed
into the base vectors of the general Hilbert space spanned over the
spatio-spectral, polarization and amplitude (field quantization) degrees of
freedom. Whereas the majority of the experiments with individual photon pairs
are realized by manipulating the states in spatio-spectral and/or polarization
degrees of freedom while keeping the state in the amplitude degree of freedom
fixed, we use the opposite configuration: We do not consider the
spatio-spectral and polarization degrees of freedom explicitly (we trace them
out) and we modify and transform the states only in the Hilbert space belonging
to the amplitude degree of freedom, i.e. the space spanned by the Fock states
of different photon numbers.

Here, we further develop and utilize the method of photon-number-resolved
postselection from TWBs to open the door for the generation of a new class of
quantum states exhibiting anticorrelations in photon-number fluctuations and
marginal sub-Poissonian statistics. To arrive at such states we consider two
TWBs with their signal beams detected together and postselection on the shared
signal beam by observing a given number of signal photons $ n_s $. The
remaining two idler beams are left in a state that exhibits strong
anticorrelations in their photon-number fluctuations. This is in striking
contrast with the usual TWBs exhibiting perfect correlations in photon numbers
as well as their fluctuations. Moreover, whereas the marginal photon-number
statistics of TWBs are super-Poissonian, the obtained states exhibit the
marginal sub-Poissonian photon-number statistics \cite{Iskhakov2016a}. These
states are prospective for metrology: They allow to measure two-photon
absorption cross-sections with the precision below the shot-noise limit, in
close analogy with the sub-shot-noise measurement of single-photon absorption
cross-section performed with a sub-Poissonian light source
\cite{Jakeman1986,Li2018,SabinesChesterkind2019}.

We note that there exists an analogy between the anticorrelations in the
photon-number fluctuations of the analyzed fields and the spatial and temporal
behavior of correlations between the signal and idler photons from a common
photon pair. Thought both photons from a photon pair usually show strong
temporal correlations \cite{Hong1986}, these correlations can be transformed
into temporal anticorrelations \cite{PerinaJr2007b}. Similarly, whereas the
signal and idler photons of usual TWBs are bunched inside their correlated
areas, there also exist the TWBs exhibiting spatial antibunching of the signal
and idler photons \cite{Nogueira2001,Caetano2003}.

The suggested scheme resembles that of the entanglement swapping suggested
first for the states of two entangled photon pairs
\cite{Zukowski1993,Scherer2009} originating in parametric down-conversion and
later also applied to swap the entanglement to the state of particles and their
collective modes \cite{Duan2001,Chou2005}. However, sensitivity of the detected
overall signal beam to the relative phase of the constituting signal beams
would be needed to observe the transfer of entanglement from the original TWBs
to the postselected idler beams. As the used TWBs are multi-mode, they are not
suitable for the entanglement transfer. Instead, in the performed experiment,
the postselection induces classical anticorrelations in photon-number
fluctuations.

To demonstrate the essence of our approach, we restrict for a moment our
attention to the states describing single-mode idler beams and consider an
ideal detector with $ n_{\rm s} $ detected signal photons (photocounts). We
model the experimental multi-mode idler beams by an incoherent superposition of
the Fock states whose statistical operator $ \hat{\varrho}_{\rm ii} $ is
written as
\begin{equation}  
 \hat{\varrho}_{\rm ii} = \sum_{{\rm i}_1=0}^{n_{\rm s}} |\alpha_{{\rm i}_1}|^2 |n_{{\rm i}_1}\rangle_{{\rm i}_1}
  {}_{{\rm i}_1}\langle n_{{\rm i}_1}|  |n_{\rm s}-n_{{\rm i}_1}\rangle_{{\rm i}_2}
  {}_{{\rm i}_2} \langle n_{\rm s}-n_{{\rm i}_1}|.
\label{1}
\end{equation}
In Eq.~(\ref{1}), a Fock state $ |n_{\rm i}\rangle_{\rm i} $ has $ n_{\rm i} $
photons in beam $ {\rm i} $ and $ \alpha_{\rm i} $ are complex coefficients.
Anticorrelations in photon-number fluctuations $ \Delta n_{\rm i} \equiv n_{\rm
i} - \langle n_{\rm i}\rangle $ represent the most striking feature of the
state $ \hat{\varrho}_{\rm ii} $. Detailed analysis reveals that even the
marginal idler-beam distributions of the analyzed states are sub-Poissonian
under suitable conditions. To understand this, let us consider for a moment the
experiment in which we independently detect the numbers of signal photons in
both signal beams. For the fixed detected signal photon numbers, both
postselected idler beams have apparently sub-Poissonian statistics. The
summation of two signal photon numbers keeping their sum fixed, as described in
Eq.~(\ref{1}), blurs the original sub-Poissonian statistics but it also
increases the success probability of the postselection process. For TWBs with
greater photon numbers and corresponding signal postselecting photon numbers
[see Fig.~\ref{fig3}(b) below], the blurring of the idler-beams photon
statistics is weak, but the success probability increases considerably. Such
states are then suitable for monitoring two-photon absorption processes or
making two-photon excitations of electronic systems.

The paper is organized as follows. The performed experiment and analysis of the
experimental data are described in Sec.~II. Sec.~III is devoted to the analysis
of the fields generated by postselection with the real detector. The properties
of the fields obtained by postselection with an ideal detector are discussed in
Sec.~IV. Detailed analysis of nonclassical properties of typical postselected
fields is contained in Sec.~V. Sec.~VI gives the conclusions. In Appendix A, a
method for fitting the experimental data with a suitable multi-mode Gaussian
field is presented. Iteration formulas for the maximum-likelihood
reconstruction are given in Appendix~B. Nonclassicality identifiers are
introduced in Appendix~C. The formula for reconstructing quasi-distributions of
integrated intensities is given in Appendix~D.

\section{Experimental setup, reconstruction and nonclassicality analysis}

To analyze the performed experiment, we consider two multi-mode and noisy TWBs
whose common mixed state is characterized by a 3D photon-number distribution $
p(n_{\rm s},n_{{\rm i}_1},n_{{\rm i}_2}) $ that gives the probability of
simultaneous presence of $ n_{\rm s} $ photons in the signal beam, $ n_{{\rm
i}_1} $ photons in the first idler beam and $ n_{{\rm i}_2} $ photons in the
second idler beam (for specific photon-number distributions, see Appendix A).
Characterizing a photon-number-resolving detector (PNRD) in the signal beam by
its detection matrix $ T_{\rm  s}(c_{\rm s},n_{\rm s}) $, that gives the
probability of detecting $ c_{\rm s} $ photocounts out of $ n_{\rm s} $
impinging photons (for details, see Appendix A), 2D photon-number distribution
$ p_{\rm ii}(n_{{\rm i}_1},n_{{\rm i}_2};c_{\rm s}) $ of a common state of the
idler beams emerging after detecting $ c_{\rm s} $ signal photocounts is
written as \cite{Saleh1978}:
\begin{equation}  
 p_{ii}(n_{{\rm i}_1},n_{{\rm i}_2};c_{\rm s}) = \sum_{n_{\rm s}=0}^{\infty}
 T_{\rm s}(c_{\rm s},n_{\rm s})p(n_{\rm s},n_{{\rm i}_1},n_{{\rm i}_2}).
\label{2}
\end{equation}
In the experiment, the postselected fields are monitored by two additional
PNRDs that give rise, together with the PNRD in the signal beam, to the 3D
experimental photocount histogram $ f(c_{\rm s},c_{{\rm i}_1},c_{{\rm i}_2}) $
that contains all information about the prepared and analyzed fields. In the
model, this histogram $ f $, as a function of the photocount numbers $ c_{\rm
s} $, $ c_{{\rm i}_1} $ and $ c_{{\rm i}_2} $ registered by three used PNRDs,
is determined along the formula
\begin{eqnarray}  
 f(c_{\rm s},c_{{\rm i}_1},c_{{\rm i}_2}) &=& \sum_{n_{\rm s}=0}^{\infty}
 T_{\rm s}(c_{\rm s},n_{\rm s}) \sum_{n_{{\rm i}_1}=0}^{\infty}
 T_{{\rm i}_1}(c_{{\rm i}_1},n_{{\rm i}_1}) \nonumber \\
 & & \times \sum_{n_{{\rm i}_2}=0}^{\infty}
 T_{{\rm i}_2}(c_{{\rm i}_2},n_{{\rm i}_2}) p(n_{\rm s},n_{{\rm i}_1},n_{{\rm i}_2})
\label{3}
\end{eqnarray}
in which the detection matrix $ T_{{\rm i}_1} $ ($ T_{{\rm i}_2} $) belongs to
the PNRD placed in the first (second) idler beam.

The reconstruction methods allow us to reveal both the conditional 2D
photon-number distributions $ p_{\rm ii} $ in Eq.~(\ref{3}) as well as the
original 3D photon-number distribution $ p $. Both a physically-motivated
method that provides a suitable Gaussian fit of the original two TWBs (see
Appendix A) and a method exploiting the maximum-likelihood approach (see
Appendix B) were applied to reconstruct the experimental photocount histogram $
f(c_{\rm s},c_{{\rm i}_1},c_{{\rm i}_2}) $ as well as the conditional
photocount histograms $ f_{\rm ii}(c_{{\rm i}_1},c_{{\rm i}_2};c_{\rm s}) $
characterizing the conditional 2D photon-number distributions $ p_{\rm
ii}(n_{{\rm i}_1},n_{{\rm i}_2};c_{\rm s}) $.

\begin{figure}  
 \centerline{\includegraphics[width=0.98\hsize]{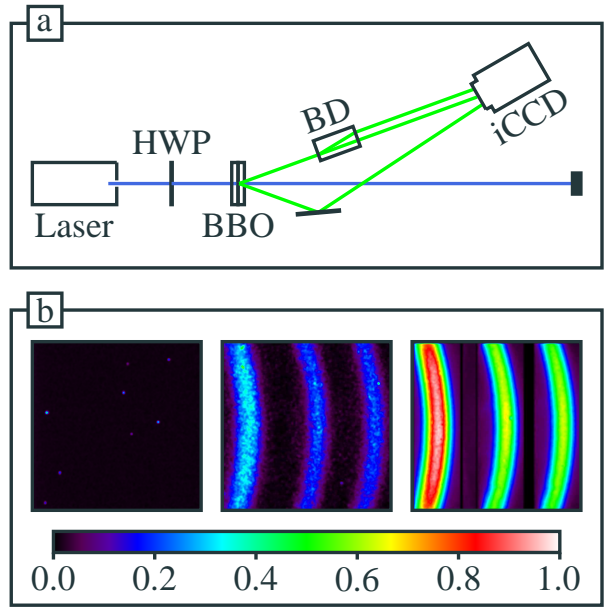}}
 \caption{(a) Scheme of the experimental setup: Laser: frequency-tripled pump laser
  with power stabilizer; HWP: half-wave plate; BBO: two thin optically contacted BBO crystals;
  BD: polarizing beam displacer; iCCD: intensified CCD camera.
  (b) Images acquired by the detector, in turn: typical single-shot image,
  accumulated image from multiple single-shot exposures forming one signal
  (left) and two idler (right) intense strips, and cumulative frame formed by individual
  detection events identified by signal processing within regions defined by one
  signal, one narrow noise (formed solely by the
  dark, ambient and readout noise) and two idler beams.}
\label{fig1}
\end{figure}

The analyzed states were prepared in the lab in the experiment whose scheme is
shown in Fig.~\ref{fig1}. Two TWBs were generated independently in type-I
spontaneous parametric down-conversion in two optically contacted 1-mm-long $
\beta $-barium-borate composite crystals (BaB$ {}_2 $O$ {}_4 $, BBO) cut for a
slightly non-collinear geometry. Whereas the first crystal gave the signal and
idler beams with horizontal polarizations, the second crystal emitted the
signal and idler beams with vertical polarizations, as a consequence of its
rotation by 90 degrees along the pump-beam propagating direction with respect
to the first crystal. Parametric down-conversion was pumped by pulses
originating in the third harmonic (280~nm) of a femtosecond cavity-dumped
Ti:sapphire laser (pulse duration 180~fs at the central wavelength of 840~nm,
repetition rate 50~kHz, pulse energy 20~nJ at the output of the third harmonic
generator). The polarization of the pump was then rotated by a half-wave plate
to balance the mutually orthogonal contributions from both crystals. The idler
beams of two TWBs that differ by their polarizations were spatially separated
by a calcite beam displacer. The signal, two idler and external noise beams
were detected in four different detection regions (in the form of strips) on
the photocathode of an iCCD camera Andor DH345-18U-63 [see the rightmost image
in Fig.~\ref{fig1}(b)]. The signal beams emitted from different crystals
spatially overlapped at the photocathode and so they were detected in a common
detection region. The camera set for 7~ns-long detection window was driven by
the synchronization electronic pulses from the laser and it operated roughly at
14~Hz frame rate. The photons of all four beams impinging on the camera were
filtered by a 14-nm-wide bandpass interference filter with the central
wavelength at 560~nm. The pump intensity, and thus also the TWBs intensity, was
actively stabilized by means of a motorized half-wave plate followed by a
polarizer and a detector that monitored the actual pump intensity.

The Gaussian reconstruction applied to the experimental photocount histogram $
f(c_{\rm s},c_{{\rm i}_1},c_{{\rm i}_2}) $ obtained after $ 1.2 \times 10^{6} $
measurement repetitions provided the following parameters for the optical
fields beyond the nonlinear crystals: The overall field was composed of two
ideal TWBs with $ 6.15 \pm 0.05 $ and $ 5.95 \pm 0.05 $ mean photon pairs ($
B_{{\rm p}_1} = 0.106 \pm 0.001 $, $ B_{{\rm p}_2} = 0.117 \pm 0.001 $, $
M_{{\rm p}_1} = 58 \pm 1 $, $ M_{{\rm p}_2} = 51 \pm 1 $) and three noise
fields with $ 0.11\pm 0.02 $, $ 0.07 \pm 0.01 $ and $ 0.02 \pm 0.01 $ mean
noise photons ($ B_{\rm s} = 10 \pm 1 $, $ B_{{\rm i}_1} = 10 \pm 1 $, $
B_{{\rm i}_2} = 39 \pm 4 $, $ M_{\rm s} = 0.011 \pm 0.001 $, $ M_{{\rm i}_1} =
0.007 \pm 0.001 $, $ M_{{\rm i}_2} = 0.0005 \pm 0.0001 $); $M_j$ stands for the
number of modes in beam $j$ having $B_j$ mean photons (photon pairs) per mode
(see Appendix~A for more details). The signal field was detected with detection
efficiency $ \eta_{\rm s} = 22.0 \pm 0.5 $~\%, detection efficiency $ \eta_{\rm
i} = 20.7 \pm 0.5 $~\% was assigned to both idler-field detection strips (lower
than the signal one due to the presence of the beam displacer). Each detection
strip was composed of $ N_{\rm s} = N_{{\rm i}_1} = N_{{\rm i}_2} = 4410 $
macropixels (one macropixel emerged from $8 \times 8$ hardware binning at the
CCD chip) and suffered from $ d_{\rm s} = d_{{\rm i}_1} = d_{{\rm i}_2} = 0.22
\pm 0.02 $ mean noise counts per detection window.

The properties of the conditional states characterized by 2D photocount [$
f_{\rm ii}(c_{{\rm i}_1},c_{{\rm i}_2};c_{\rm s}) $] and photon-number
distributions [$ p_{\rm ii}(n_{{\rm i}_1},n_{{\rm i}_2};c_{\rm s}) $] were
quantified by the following parameters. Anticorrelation between the
fluctuations $ \Delta n $ ($ \Delta c $) of the idler-fields photon
(photocount) numbers was recognized by negative values of the covariance $
C_{n,\Delta} $,
\begin{equation}  
 C_{n,\Delta} = \frac{ \langle \Delta n_{{\rm i}_1}\Delta n_{{\rm i}_2}\rangle }{
  \sqrt{ \langle (\Delta n_{{\rm i}_1})^2\rangle \langle(\Delta n_{{\rm i}_2})^2\rangle }
  }.
\label{4}
\end{equation}
Nonclassical character of the conditional 2D idler fields is verified by the
values of the modified noise-reduction-parameter $ R_{n,+} $ smaller than 1,
\begin{equation}  
 R_{n,+} = \frac{ \langle (\Delta (n_{{\rm i}_1}+n_{{\rm i}_2})^2\rangle }{
   \langle n_{{\rm i}_1}\rangle + \langle n_{{\rm i}_2} \rangle }.
\label{5}
\end{equation}
We have $ R_{n,+} = 1 $ for two independent Poissonian fields in coherent
states. Declinations of classical photon-number distributions from the
Poissonian ones as well as mutual photon-number correlations between the fields
increase the values of the modified noise-reduction-parameter $ R_{n,+} >1 $.
On the other hand, the inequality $ R_{n,+} < 1 $ is equivalent to the
inequality for the moments of integrated intensities $ \langle [\Delta(W_{{\rm
i}_1}+W_{{\rm i}_2})]^2\rangle \equiv \int_0^\infty dW_{{\rm i}_1}
\int_0^\infty dW_{{\rm i}_2} [\Delta(W_{{\rm i}_1}+W_{{\rm i}_2})]^2 P_{\cal
N}(W_{{\rm i}_1},W_{{\rm i}_2}) < 0 $. Its fulfillment requires the
quasi-distribution $ P_{\cal N}(W_{{\rm i}_1},W_{{\rm i}_2}) $ of integrated
intensities with negative values which implies the fields' nonclassicality. We
note that the integrated intensities $ W $ and their moments occur in the
description of optical fields in relation to their detection as the fields
detectors are sensitive to the normally-ordered photon-number moments that are
referred to as the moments of integrated intensity [for the relation between
both types of moments, see Eq.~(\ref{A6}) in Appendix~A]. We have $ R_{n,+} = 0
$ for the state in Eq.~(\ref{1}). Thus, this state in nonclassical. On the
other hand, it is not entangled as it contains only classical anticorrelations
in photon-number fluctuations.

Also the marginal idler fields may exhibit the nonclassical sub-Poissonian
statistics observed when the values of the Fano factors $ F_{n,{{\rm i}_j}} $,
$ j=1,2 $,
\begin{equation}  
 F_{n,{{\rm i}_j}} = \frac{ \langle (\Delta n_{{\rm i}_j})^2\rangle }{
   \langle \Delta n_{{\rm i}_j}\rangle } ,
\label{6}
\end{equation}
are smaller than 1.

The nonclassicality of conditional 2D idler fields may be identified both using
the nonclassicality criteria (NCCa) written in terms of the intensity moments
and probabilities of photon-number (photocount) distributions. The NCCa using
the intensity moments $ C_W $ and~$ M_W $,
\begin{eqnarray}  
 C_W &\equiv& \langle W_{{\rm i}_1}^{2}W_{{\rm i}_2}^{2}\rangle
  - \langle W_{{\rm i}_1}W_{{\rm i}_2}\rangle^2 < 0, \label{7} \\
 M_W &\equiv& \langle W_{{\rm i}_1}^{2}\rangle\langle W_{{\rm i}_2}^{2}\rangle
   + 2\langle W_{{\rm i}_1}W_{{\rm i}_2}\rangle\langle W_{{\rm i}_1}
   \rangle\langle W_{{\rm i}_2}\rangle
   - \langle W_{{\rm i}_1}W_{{\rm i}_2}\rangle^2 \nonumber \\
  & & \mbox{} - \langle W_{{\rm i}_1}^2\rangle\langle W_{{\rm i}_2}\rangle^2
   - \langle W_{{\rm i}_1}\rangle^2\langle W_{{\rm i}_2}^2\rangle < 0.
   \label{8}
\end{eqnarray}
derived from the Cauchy--Schwarz inequality and the matrix approach
\cite{PerinaJr2020}, respectively, have been found the most powerful for the
analyzed states. They belong to the groups of the NCCa discussed in Appendix C
[$ C_W = C_{(1,1)}^{(2,0)} $, $ M_W = M_{(0,0),(1,0),(0,1)} $]. Their
probability variants are then used to identify the location of nonclassicality
across the probability distributions.

Sub-Poissonian character of the marginal idler fields makes the following
hybrid NCC $ L $ \cite{Arkhipov2016c} very efficient in revealing the
nonclassicality:
\begin{equation}   
  L_{Wp}(n_{{\rm i}_1}) \equiv \langle W_{i_2}^{3}\rangle_{n_{{\rm i}_1}} \langle W_{i_2}\rangle_{n_{{\rm i}_1}} -
    \langle W_{i_2}^{2}\rangle_{n_{{\rm i}_1}}^2  <0.
\label{9}
\end{equation}
In Eq.~(\ref{9}), averaging $ \langle \rangle_{n_{{\rm i}_1}} $ is performed in
the variable $ n_{{\rm i}_2} $ with the photon-number distribution $ p(n_{{\rm
i}_1},n_{{\rm i}_2}) $ in which $ n_{{\rm i}_1} $ is kept fixed. This means
that the intensity moments are determined in one variable whereas the
probabilities are used in the other to reveal the nonclassicality.

When applying the concept of the Lee nonclassicality depth (NCD) \cite{Lee1991}
the NCCa also provide quantification of the nonclassicality. The NCD $ \tau $
is derived from the value $ s_{\rm th} $ of the ordering parameter at which the
used NCC loses its ability to reveal the nonclassicality of the analyzed field
\cite{PerinaJr2020}:
\begin{equation}  
  \tau = (1- s_{\rm th})/2 .
\label{10}
\end{equation}
To determine the threshold values $ s_{\rm th} $, transformations of the
photon-number distributions as well as the intensity moments between different
field's orderings are needed \cite{Perina1991,PerinaJr2017a,PerinaJr2020a}.

\section{Nonclassical light generated by postselection with the real detector}

\begin{figure}  
 \includegraphics[width=0.48\hsize]{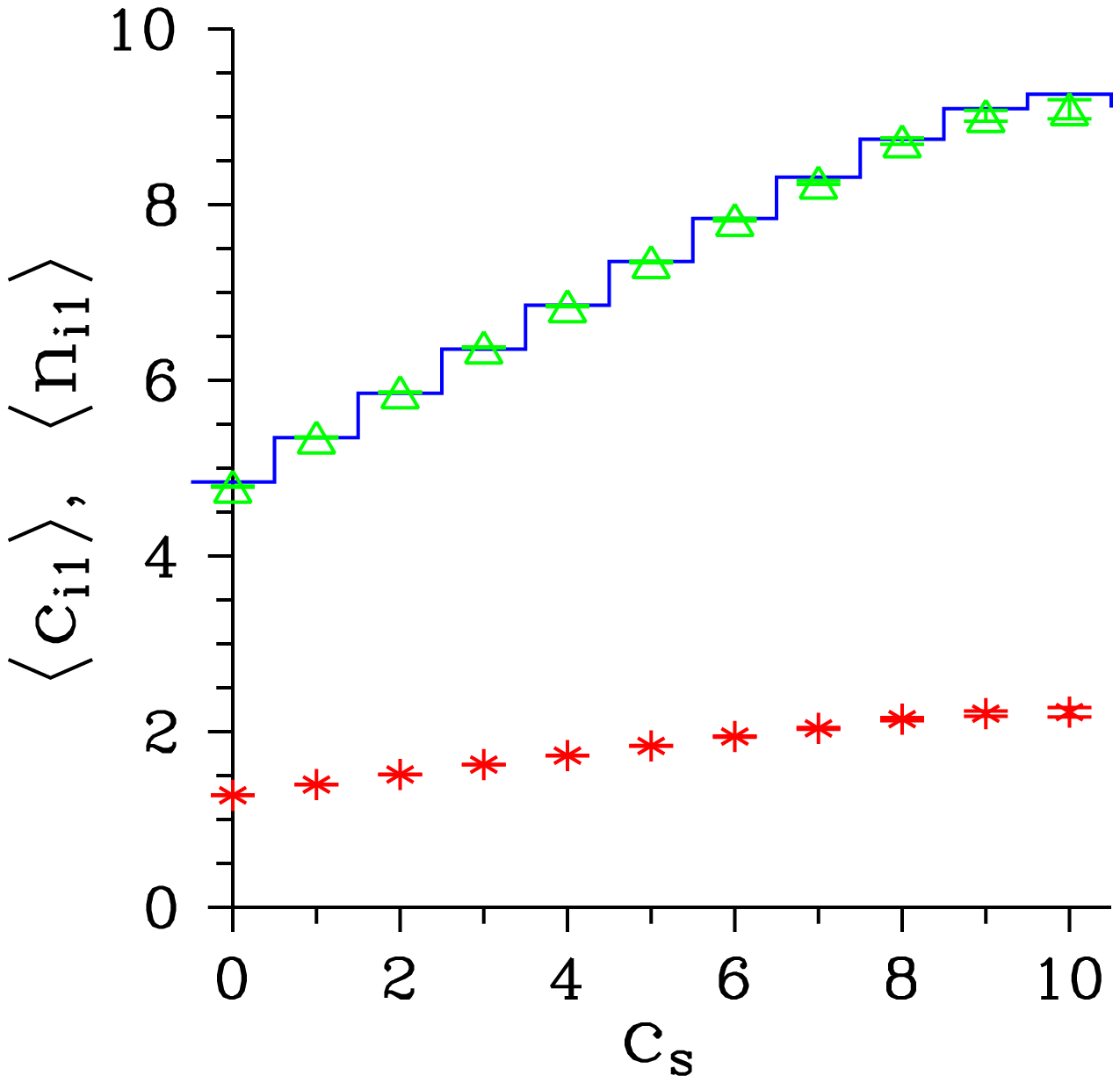}
 \includegraphics[width=0.48\hsize]{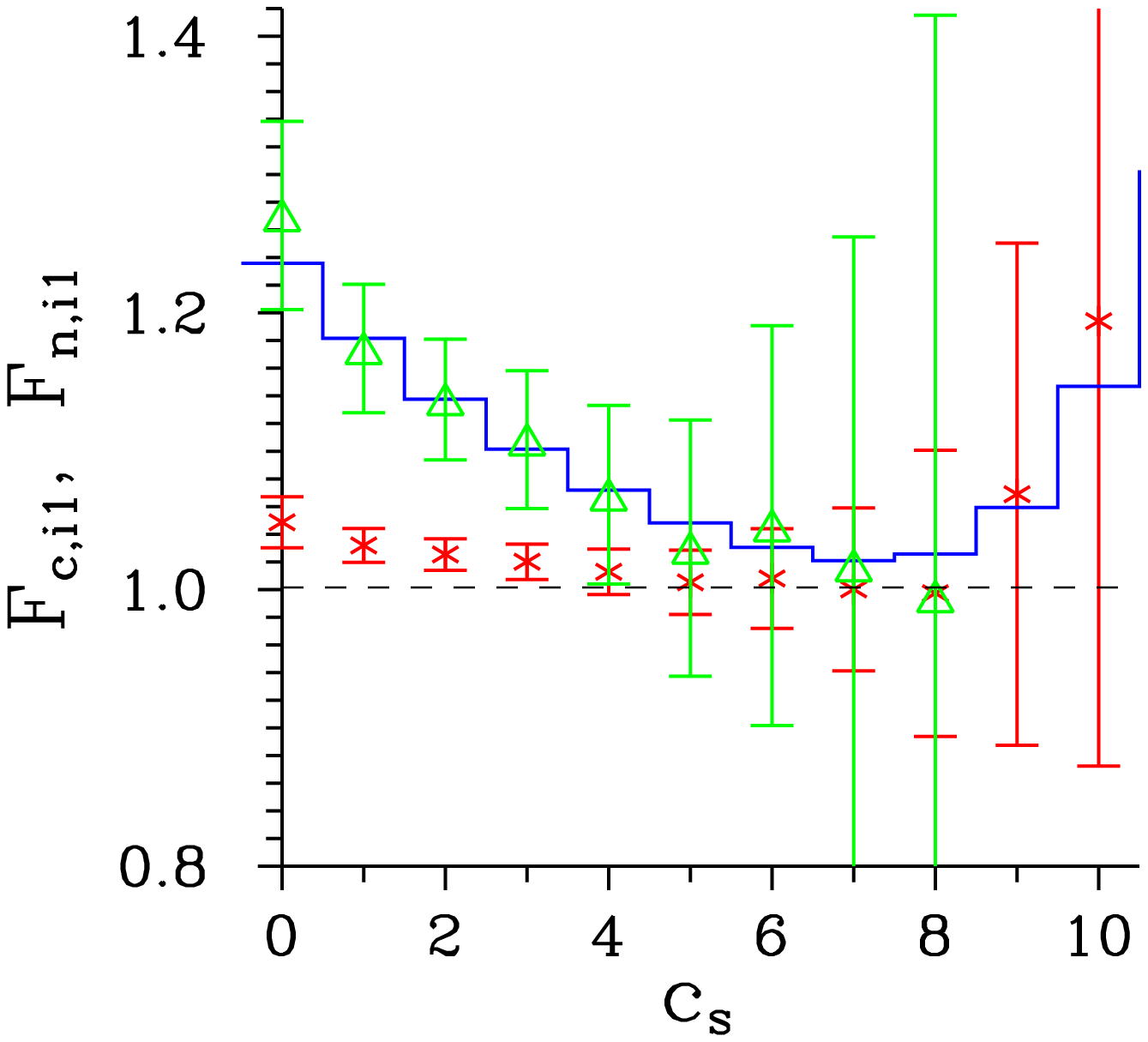}
  \vspace{2mm}
  \centerline{ \small (a) \hspace{.45\hsize} (b)}

 \includegraphics[width=0.48\hsize]{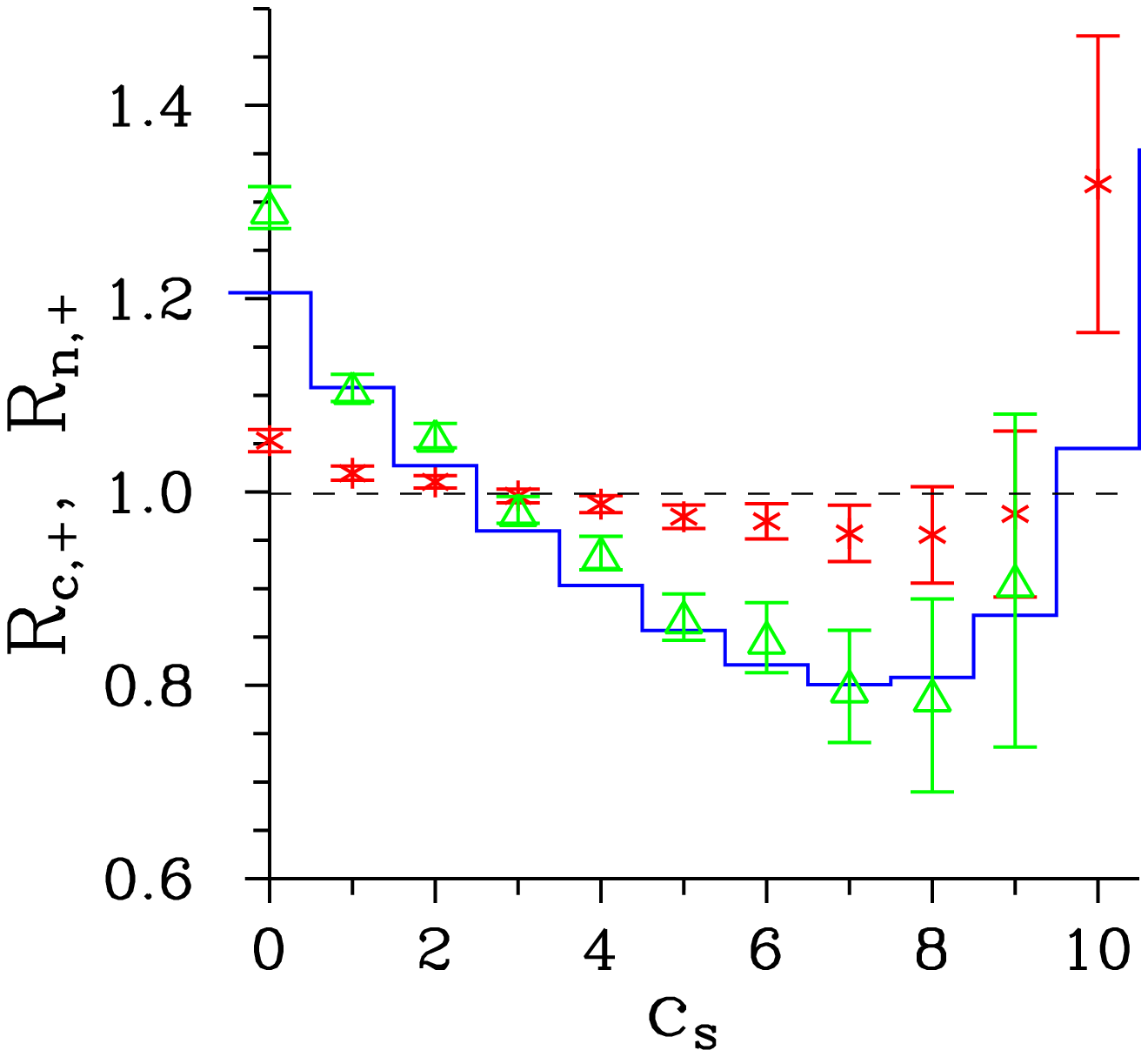}
 \includegraphics[width=0.48\hsize]{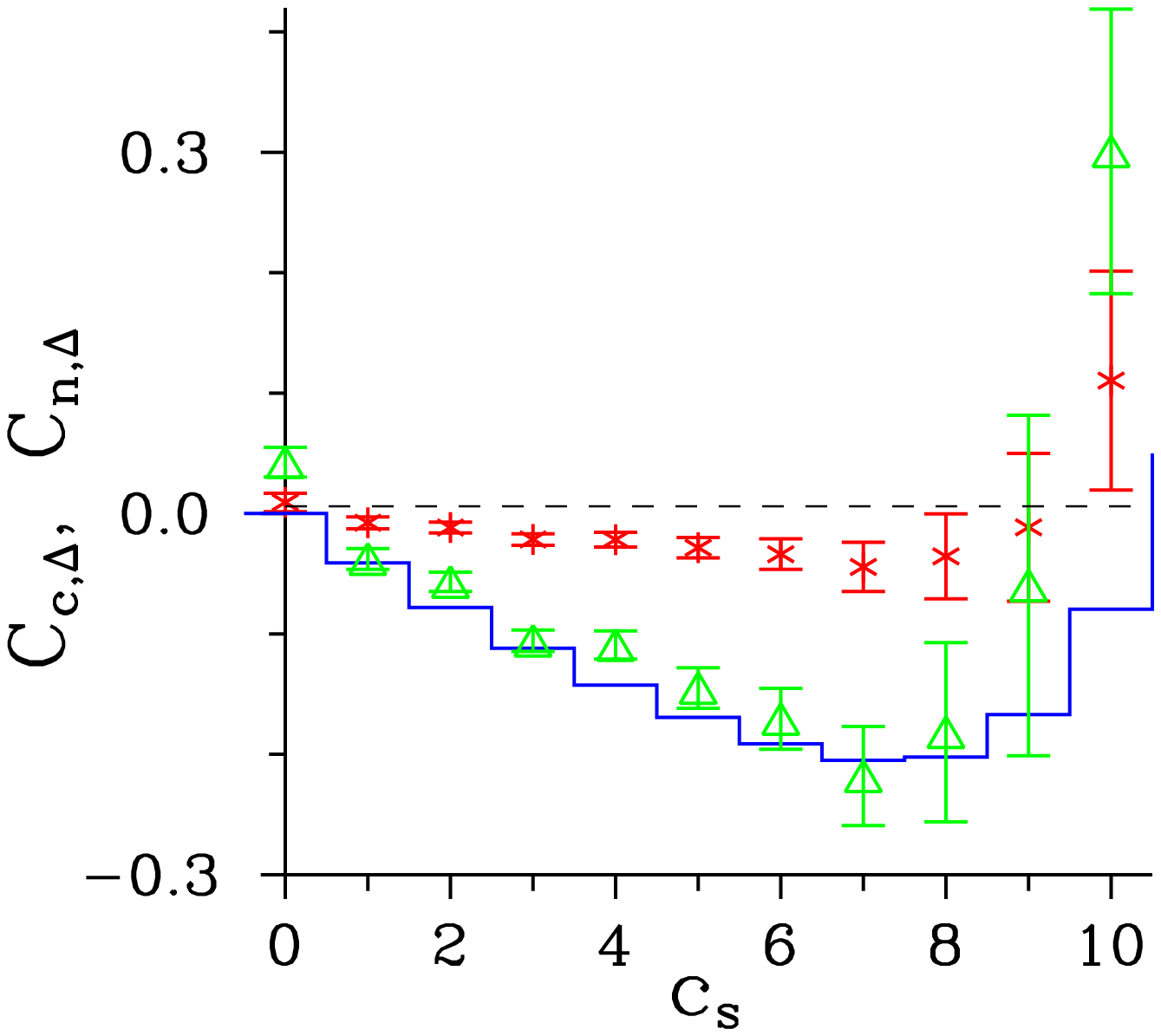}
  \vspace{2mm}
  \centerline{ \small (c) \hspace{.45\hsize} (d)}

 \includegraphics[width=0.48\hsize]{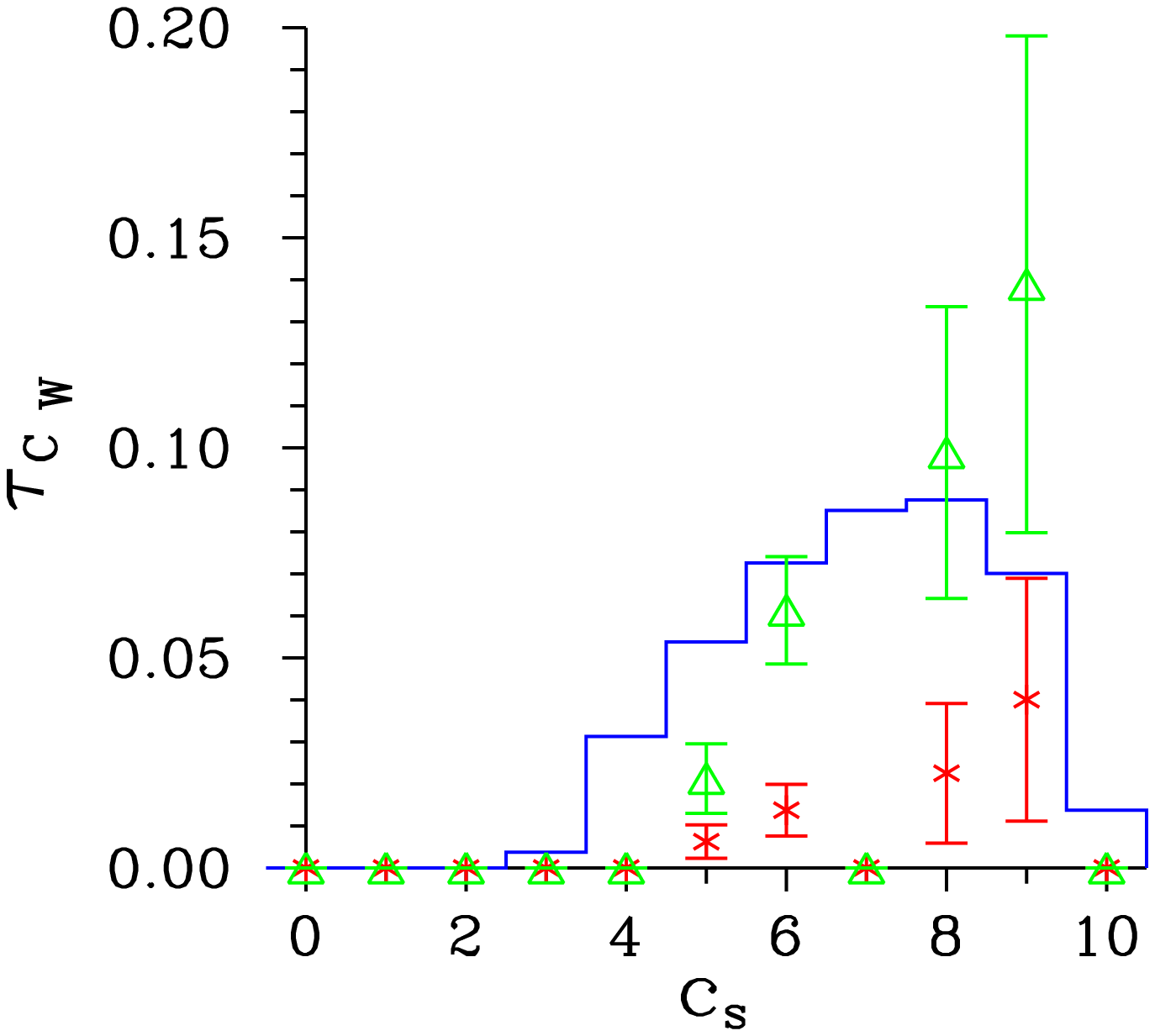}
 \includegraphics[width=0.48\hsize]{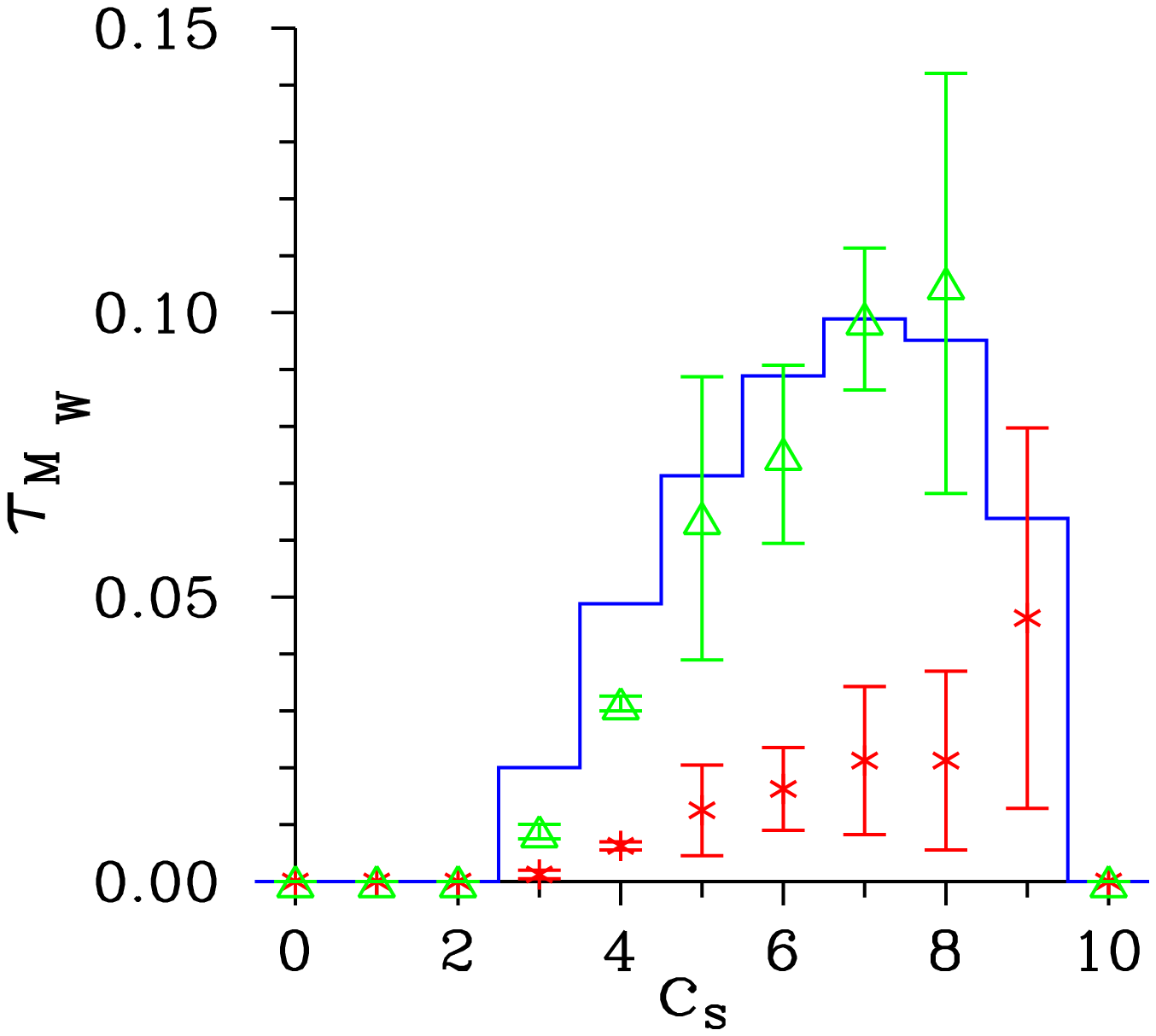}
  \vspace{2mm}
  \centerline{ \small (e) \hspace{.45\hsize} (f)}

 \caption{(a) Mean number of photons $ \langle n_{{\rm i}_1}\rangle $ (photocounts $ \langle c_{{\rm i}_1}\rangle $) and (b)
  Fano factor $ F_{n,{\rm i}_1} $ ($ F_{c,{\rm i}_1} $) of the first idler field, (c) modified noise-reduction-parameter $ R_{n,+} $ ($ R_{c,+} $),
  (d) covariance $ C_{n,\Delta} $ ($ C_{c,\Delta} $), and
  nonclassicality depths (e) $ \tau_{C_W} $ and (f) $ \tau_{M_W} $ of the 2D
  idler fields observed after postselection as they depend on the signal-field postselecting photocount
  number $ c_{\rm s} $. Isolated symbols are drawn for the experimental photocount
  histograms (red $ \ast $) and fields reconstructed by 2D maximum-likelihood
  approach (green $ \triangle $); solid blue curves originate in the 3D Gaussian
  model. The horizontal dashed lines indicate the borders of anticorrelation ($ C_{\Delta} = 0 $) and nonclassicality
  ($ F_{{\rm i}_1} = 1 $, $ R_{+} = 1 $) regions.}
\label{fig2}
\end{figure}

First, we analyze the experimental 2D photocount histograms $ f_{\rm
ii}(c_{{\rm i}_1},c_{{\rm i}_2};c_{\rm s}) $ and the corresponding
reconstructed photon-number distributions reached by the maximum-likelihood
approach [$ p_{\rm ii}^{ML}(n_{{\rm i}_1},n_{{\rm i}_2};c_{\rm s}) $, see
Appendix B] and the suitable Gaussian fit [$ p_{\rm ii}^{G}(n_{{\rm
i}_1},n_{{\rm i}_2};c_{\rm s}) $, see Appendix A] from the point of view of the
marginal idler-fields mean photocount [$ \langle c_{{\rm i}_j}\rangle $, $
j=1,2 $] and photon [$ \langle n_{{\rm i}_j}\rangle $] numbers and the Fano
factors [$ F_{{\rm i}_j} $] that quantify the spread of photocount and
photon-number fluctuations. Both marginal idler fields behave similarly. The
mean photocount [$ \langle c_{{\rm i}_1}\rangle $] and photon [$ \langle
n_{{\rm i}_1}\rangle $] numbers of the first idler field increase with the
increasing postselected signal photocount number $ c_{\rm s} $ in the analyzed
range $ c_{\rm s} \le 10 $, as shown in Fig.~\ref{fig2}(a). On the other hand,
the relative fluctuations in photocount and photon numbers as quantified by the
Fano factors $ F_{c,{\rm i}_1} $ and $ F_{n,{\rm i}_1} $ in Fig.~\ref{fig2}(b)
decrease with the increasing $ c_{\rm s} $ up to $ c_{\rm s} = 7 $ and then
they increase. This is a consequence of the postselection mechanism between the
signal and the first idler field that suffers from non-unit detection
efficiency $ \eta_{\rm s} $ and the noise signal photons together with the
signal-detector dark counts. Whereas the detrimental role of non-unit detection
efficiency $ \eta_{\rm s} $ on the Fano factor $ F $ decreases with the
increasing signal photocount number $ c_{\rm s} $, the effect of the noise
signal photons and dark-counts behaves in the opposed way \cite{PerinaJr2013b}.
Also the experimental errors of the Fano factor $ F $ increase with the
increasing $ c_{\rm s} $ which is a consequence of the decreasing number of
measurement repetitions associated with a given signal photocount number $
c_{\rm s} $. Owing to the relatively low detection efficiency $ \eta_{\rm
s}\approx 20~\% $ and large relative portion of the noise in the signal field
(around 1/2 caused by the signal photons from the second TWB) the values of
Fano factor $ F $ remain in the classical region with $ F \ge 1 $.

However, when we analyze the performance of the postselection mechanism on the
sum $ c_{{\rm i}_1} + c_{{\rm i}_2} $ ($ n_{{\rm i}_1} + n_{{\rm i}_2} $) of
the first and the second idler photocount (photon) numbers, i.e. when the
postselection mechanism works simultaneously and 'in-phase' on both TWBs, we
get the reduction of fluctuations of the above sums below their classical
border ($ R_{+} < 1 $) for $ c_{\rm s} \in \langle 3,9\rangle $, as documented
by the modified noise-reduction-parameters $ R_{c,+} $ and $ R_{n,+} $ plotted
in Fig.~\ref{fig2}(c). The smallest values of $ R_{+} $ indicating the
strongest achieved suppression of the fluctuations are reached for the signal
photocount numbers $ c_{\rm s} = 7,8 $, in accordance with the behavior of the
first and the second idler-field Fano factors $ F_{{\rm i}_1} $ and $ F_{{\rm
i}_2} $. The suppression of fluctuations in the sum $ c_{{\rm i}_1} + c_{{\rm
i}_2} $ ($ n_{{\rm i}_1} + n_{{\rm i}_2} $) of the idler-fields photocount
(photon) numbers quantified by $ R_{c,+} < 1 $ ($ R_{n,+} < 1 $) gives rise to
strong anticorrelations between the fluctuations of the first and the second
photocount (photon) numbers $ \Delta c_{{\rm i}_1} $ ($ \Delta n_{{\rm i}_1} $)
and $ \Delta c_{{\rm i}_2} $ ($ \Delta n_{{\rm i}_2} $). They are alternatively
quantified by the covariances $ C_{c,\Delta} $ and $ C_{n,\Delta} $ drawn in
Fig.~\ref{fig2}(d).

Contrary to the case of TWBs, revealing the nonclassicality of the postselected
2D fields is much harder. Out of numerous NCCa written in intensity moments and
successfully applied to TWBs in \cite{PerinaJr2017a}, only the NCC $ C_W $ in
Eq.~(\ref{7}) derived from the Cauchy--Schwarz inequality and the NCC $ M_W $
in Eq.~(\ref{8}) originating in the matrix approach provided high and
comparable values of the corresponding NCDs $ \tau_{C_W} $ and $ \tau_{M_W} $,
as shown in Figs.~\ref{fig2}(e,f). The comparison of graphs in
Figs.~\ref{fig2}(e) and \ref{fig2}(f) drawn for the experimental 2D photocount
histograms and photon-number distributions provided by 2D maximum-likelihood
approach reveals the NCC $ M_W $ as more stable and reliable because it
identifies all the states postselected by detecting the signal photocount
numbers $ c_{\rm s} \in \langle 3,9\rangle $ as nonclassical, in accordance
with the values of the modified noise-reduction-parameter $ R_{n,+} $ plotted
in Fig.~\ref{fig2}(c).

We note that the classical/nonclassical features identified in the experimental
photocount histograms $ f_{\rm ii} $ are emphasized in the photon-number
distributions $ p_{\rm ii} $ obtained by both reconstruction methods, as
documented in Figs.~\ref{fig2}(b--f).

\section{Nonclassical light generated by postselection with an ideal detector}

Detection of the postselecting signal field with a better detection efficiency
$ \eta_{\rm s} $ opens the door for the observation of the postselected 2D
idler fields with their most pronounced properties: anticorrelation in the
idler-field photon-number fluctuations and sub-Poissonian statistics in the
marginal idler fields. We demonstrate these properties by reconstructing the
whole optical field as it occurs in front of all three used PNRDs, i.e. we also
involve the signal-field postselecting detector in the reconstruction. We
accomplish the reconstruction both by applying the 3D maximum-likelihood
approach (see Appendix B) and a suitable 3D Gaussian fit (see Appendix A) to
the experimental photocount histogram $ f(c_{\rm s},c_{{\rm i}_1},c_{{\rm
i}_2}) $. Then, similarly as above, we analyze the 2D idler-fields
photon-number distributions $ p(n_{{\rm i}_1},n_{{\rm i}_2};n_{\rm s}) $
conditioned by the presence of $ n_{\rm s} $ photons in the signal field. This
corresponds to the use of an ideal PNRD in the postselection mechanism.

\begin{figure}  
 \includegraphics[width=0.48\hsize]{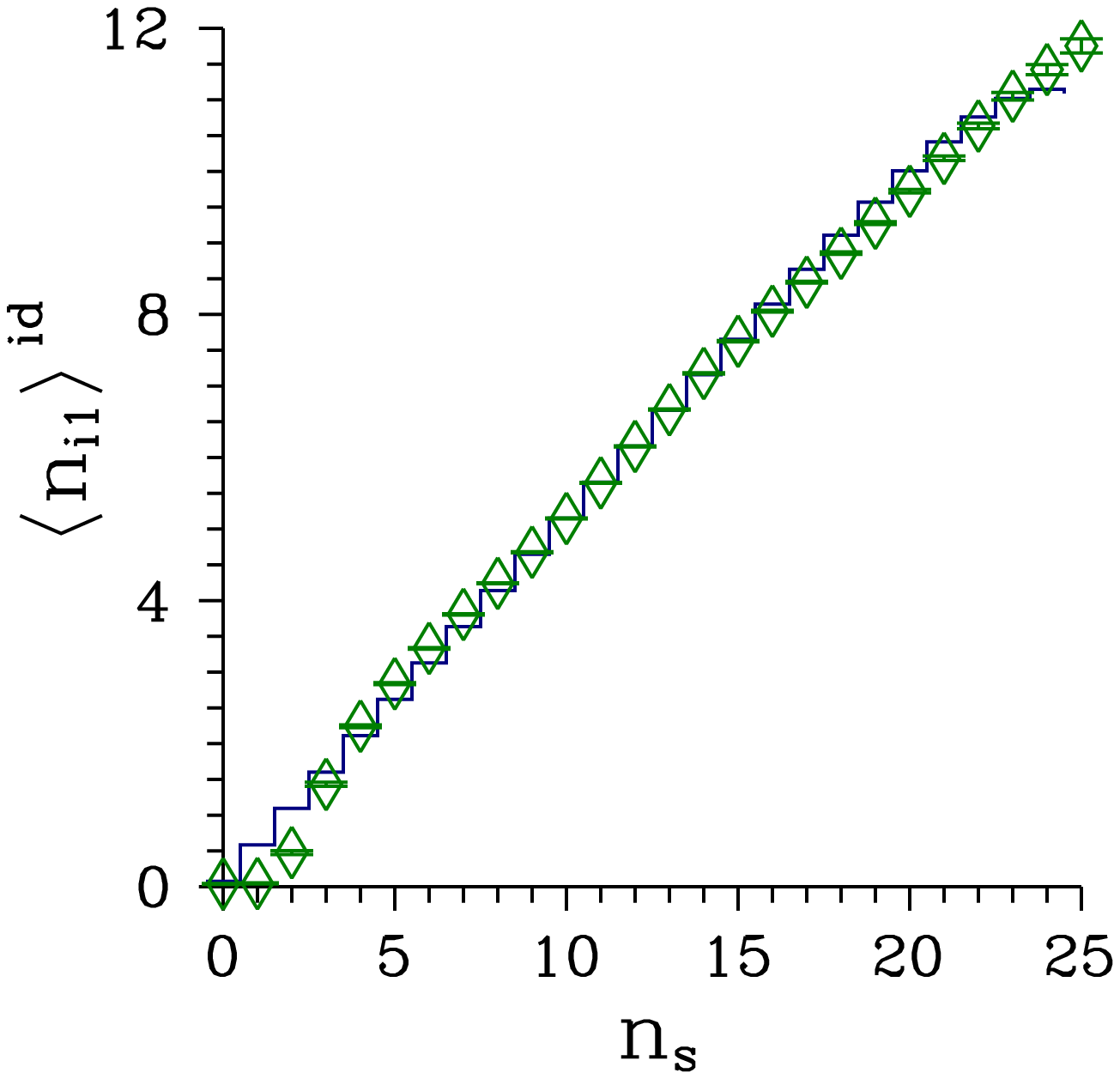}
 \includegraphics[width=0.48\hsize]{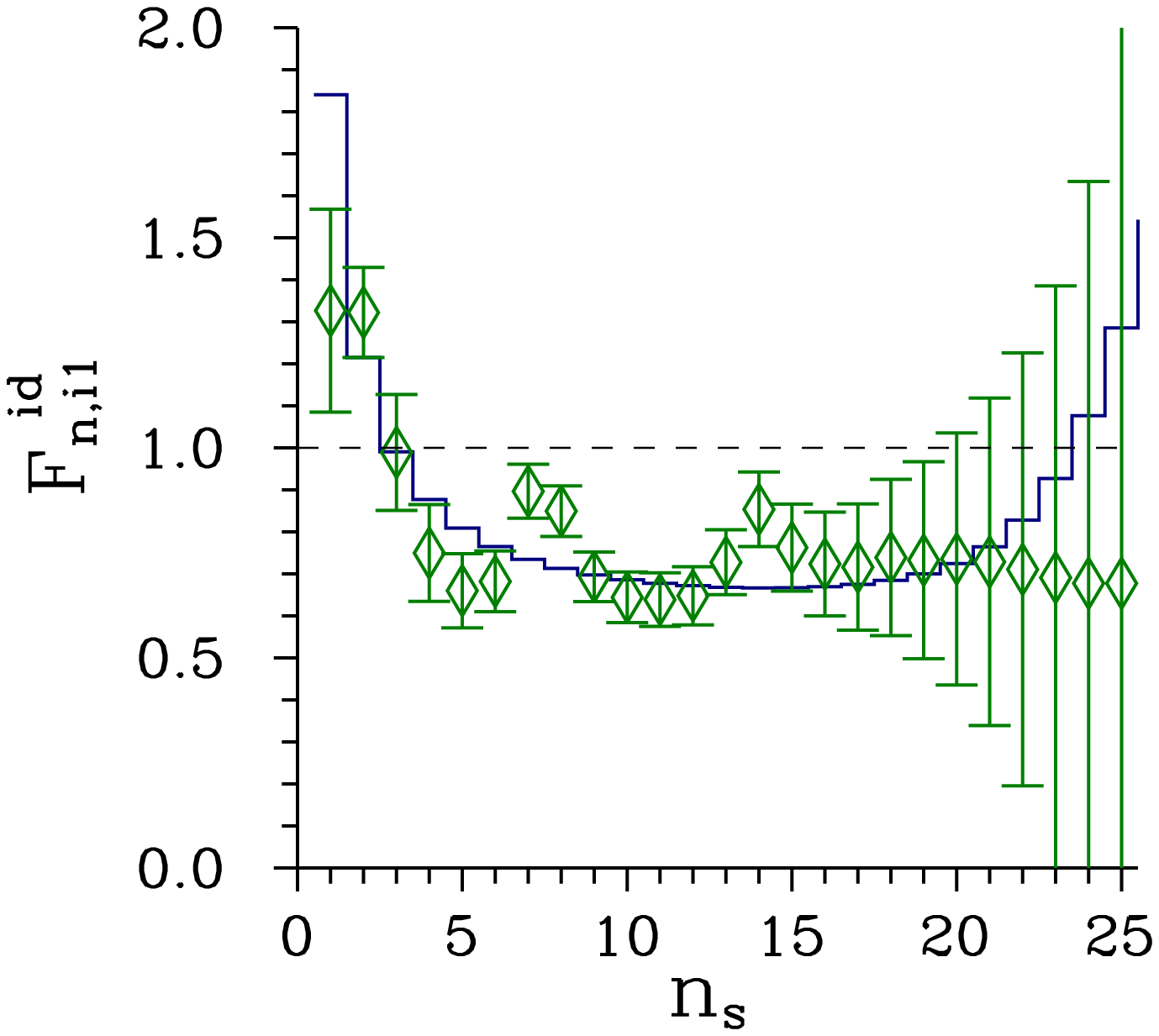}
  \vspace{2mm}
  \centerline{ \small (a) \hspace{.45\hsize} (b)}

 \includegraphics[width=0.48\hsize]{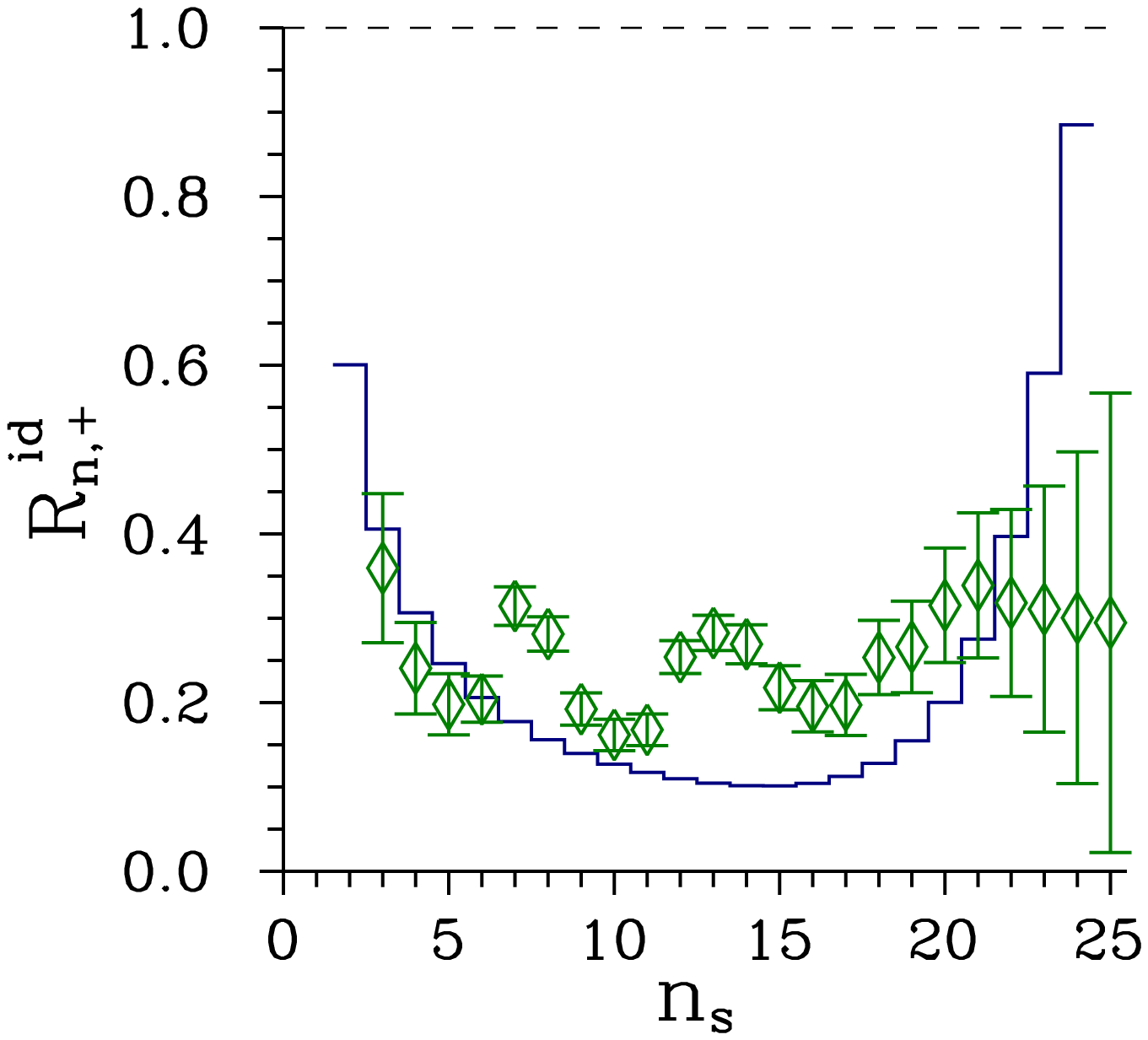}
 \includegraphics[width=0.48\hsize]{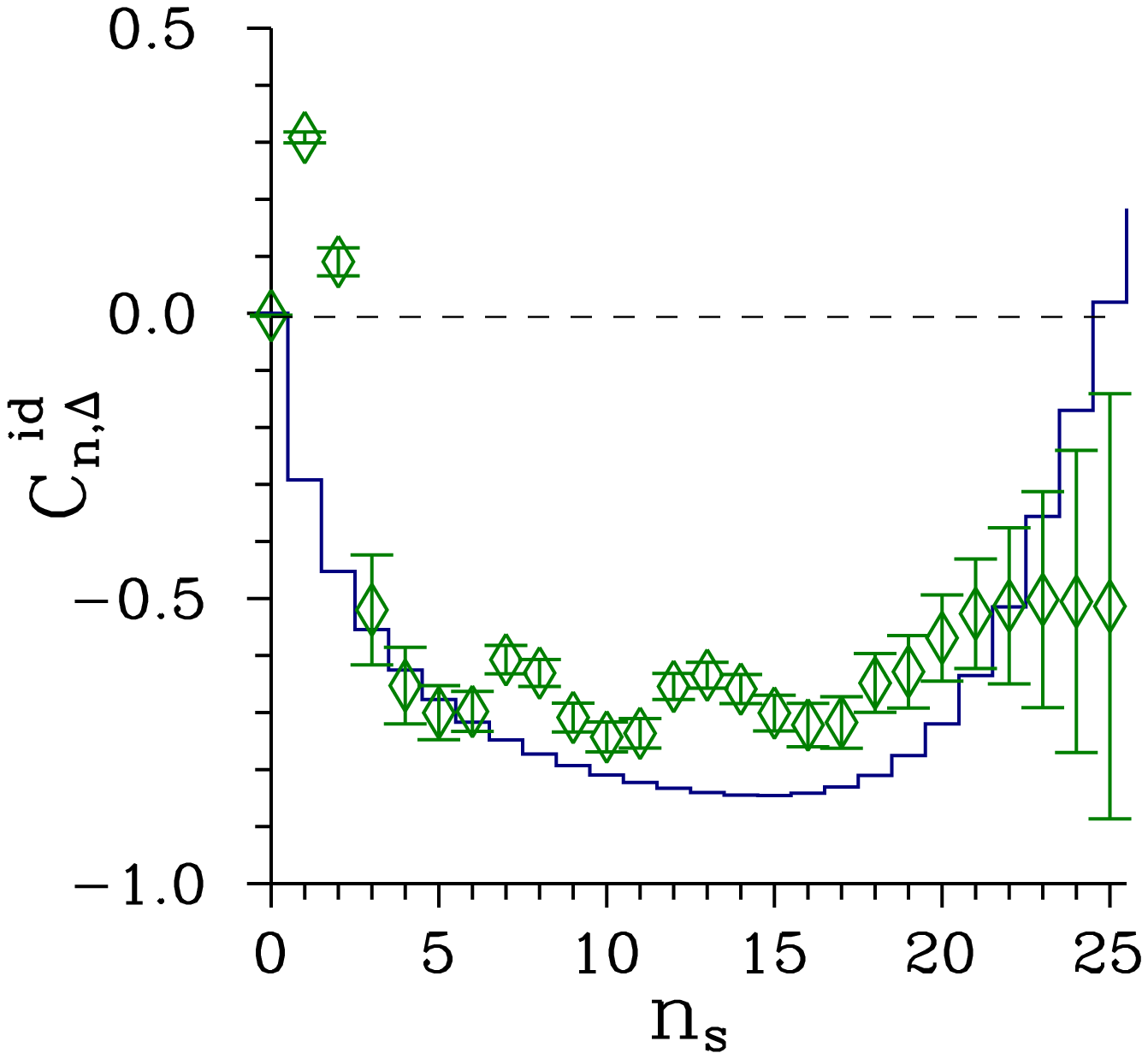}
  \vspace{2mm}
  \centerline{ \small (c) \hspace{.45\hsize} (d)}

 \includegraphics[width=0.48\hsize]{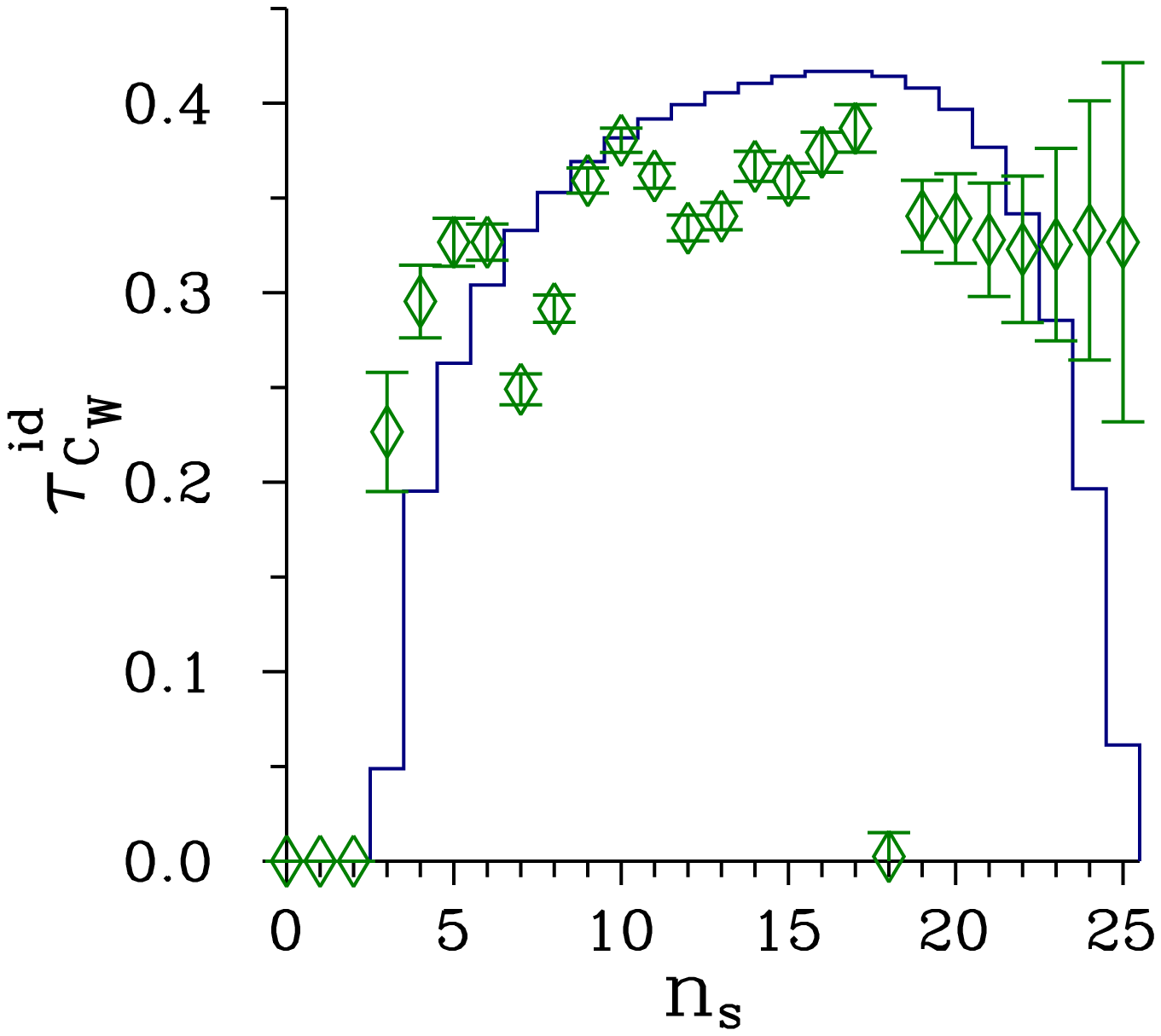}
 \includegraphics[width=0.48\hsize]{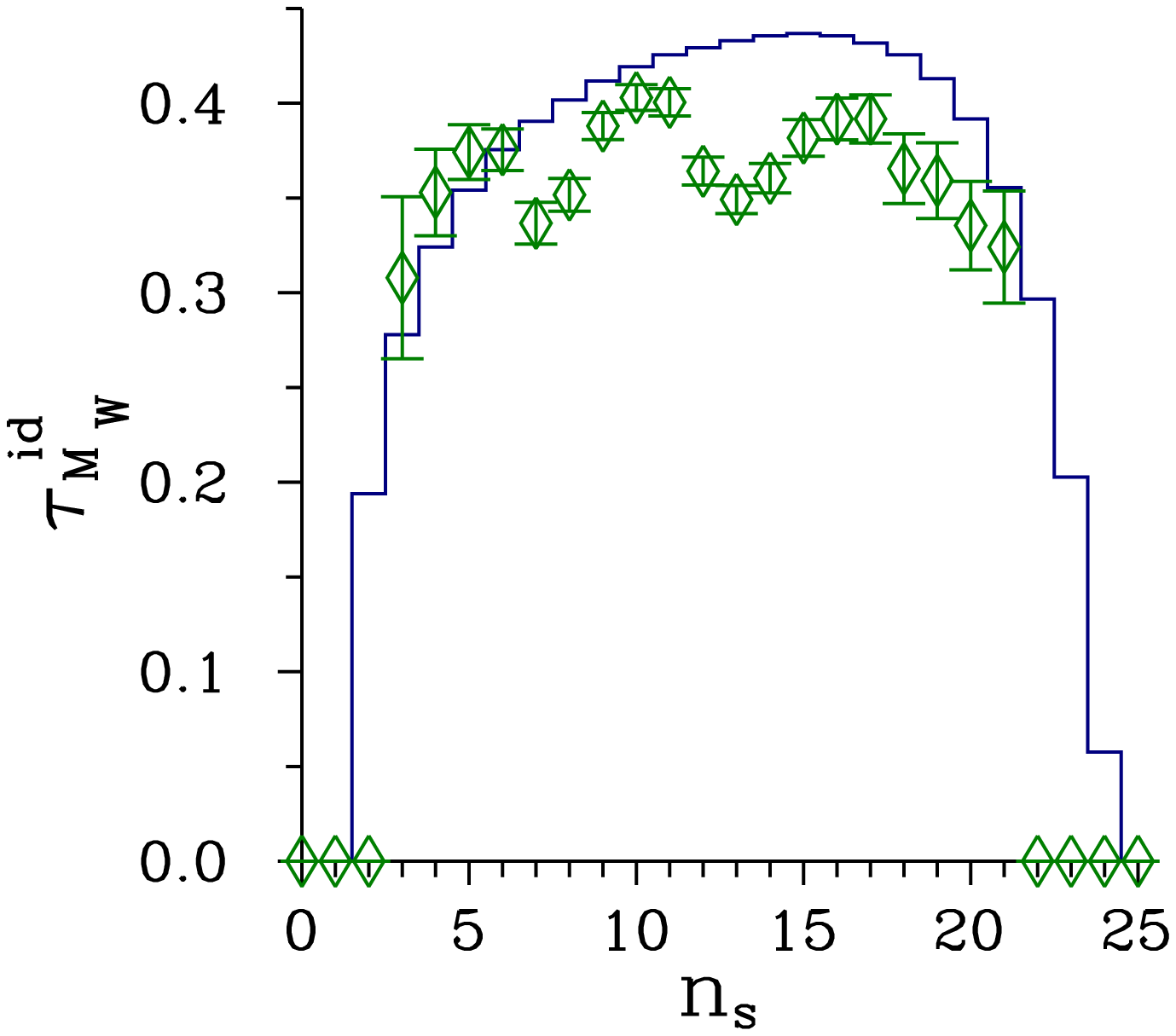}
  \vspace{2mm}
  \centerline{ \small (e) \hspace{.45\hsize} (f)}

 \caption{(a) Mean number of photons $ \langle n_{{\rm i}_1}\rangle^{\rm id} $ and (b)
  Fano factor $ F_{n,{\rm i}_1}^{\rm id} $ of the first idler field, (c) modified noise-reduction-parameter $ R_{n,+}^{\rm id} $,
  (d) covariance $ C_{n,\Delta}^{\rm id} $, and
  nonclassicality depths (e) $ \tau_{C_W}^{\rm id} $ and (f) $ \tau_{M_W}^{\rm id} $ of the 2D
  idler fields reached by the ideal photon-number-resolving postselection as they depend on the signal-field
  postselecting photon number $ n_{\rm s} $. Isolated symbols are drawn for the field reconstructed by 3D maximum-likelihood
  approach (dark green $ \diamond $); solid dark blue curves originate in the 3D Gaussian
  model. The horizontal dashed lines indicate the borders of anticorrelation ($ C_{\Delta}^{\rm id} = 0 $) and nonclassicality
  ($ F_{{\rm i}_1}^{\rm id} = 1 $, $ R_{+}^{\rm id} = 1 $) regions.}
\label{fig3}
\end{figure}

The postselected idler fields behave similarly also in this case. The mean
photon numbers $ \langle n_{{\rm i}_1}\rangle^{\rm id} $ of the first idler
field increase roughly linearly with the postselecting signal photon number $
n_{\rm s} $, and we have $ \langle n_{{\rm i}_1}\rangle^{\rm id} \approx n_{\rm
s}/2 $ [see Fig.~\ref{fig3}(a)]. Owing to the ideal detection efficiency $
\eta_{\rm s} = 1 $ the Fano factors $ F_{{\rm i}_1} $ attain nonclassical
values ($ F < 1 $) for greater signal photon numbers $ n_{\rm s} $. According
to the graph in Fig.~\ref{fig3}(b), the Fano factors $ F^{\rm id} $ smaller
than 0.7 are reached for the signal photon numbers $ n_{\rm s} \in \langle
4,20\rangle $. For the reconstructed 3D Gaussian field, sub-Poissonian
character of the marginal idler fields is lost fast for even greater values of
$ n_{\rm s} $ as a consequence of the noise signal photons originating in the
second TWB.

The sub-Poissonian Fano factors of the marginal idler fields reflect efficient
functioning of the postselection mechanism that gives raise to low values of
the modified noise-reduction-parameter $ R_{n,+}^{\rm id} $. According to
Fig.~\ref{fig3}(c) they attain the highly-nonclasical values around 0.2 --- 0.3
in the whole range $ n_{\rm s} \in \langle 4,20\rangle $. Also the covariance $
C_{n,\Delta}^{\rm id} $ of the idler-field photon-number fluctuations $ \Delta
n_{{\rm i}_1} $ and $ \Delta n_{{\rm i}_2} $ plotted in Fig.~\ref{fig3}(d)
attains the values around -0.8 --- -0.6 in this range, which expresses the
strong anticorrelation. Whereas the greatest values of the NCDs $ \tau_{C_W} $
and $ \tau_{M_W} $ reached by the real detector equal around 0.1, the
postselection by the ideal detector provides the much-greater values of up to
around 0.4, as documented in Figs.~\ref{fig3}(e,f). The comparison of graphs in
Figs.~\ref{fig3}(e) and \ref{fig3}(f) plotted for the photon-number
distributions originating in the 3D maximum-likelihood approach reveals the NCC
$ M_W $ more stable than the NCC $ C_W $ in identifying and quantifying the
nonclassicality.

In the quantities plotted in Figs. \ref{fig3}(b-f) there occur little
oscillations with the increasing period as the postselecting signal photon
number $ n_{\rm s} $ increases. They originate in the discrete photocount
numbers $ c_{\rm s} $ provided by the measurement. The 3D maximum-likelihood
reconstruction has to correct for the detection efficiency $ \eta_{\rm s}
\approx 20\% $: The neighbor measurements for $ c_{\rm s} $ and $ c_{\rm s}+1 $
postselecting photocounts have to be expanded into the interval of $ n_{\rm s}
$ postselecting photons from $ \approx c_{\rm s}/\eta_{\rm s} $ to $ \approx
c_{\rm s}/\eta_{\rm s} + 1/\eta_{\rm s} $. Gradual stretching of the
oscillation period $ \approx 5 $ is then caused by the presence of dark counts.
The oscillations reflect the varying quality of the measurement for different
postselecting photon numbers $ n_{\rm s} $: The measurements for the numbers $
n_{\rm s} $ for which $ \eta_{\rm s} n_{\rm s} $ are close to integers are of
the best quality and allow to reconstruct the studied quantities in the best
way. For the remaining numbers $ n_{\rm s} $ the measurements are, roughly
speaking, split between the neighbor photocount numbers $ c_{\rm s} $ and so
their quality as well as the quality of the reconstructed quantities are worse.

\section{Detailed analysis of nonclassical properties of postselected fields}

\begin{figure}  
 \includegraphics[width=0.48\hsize]{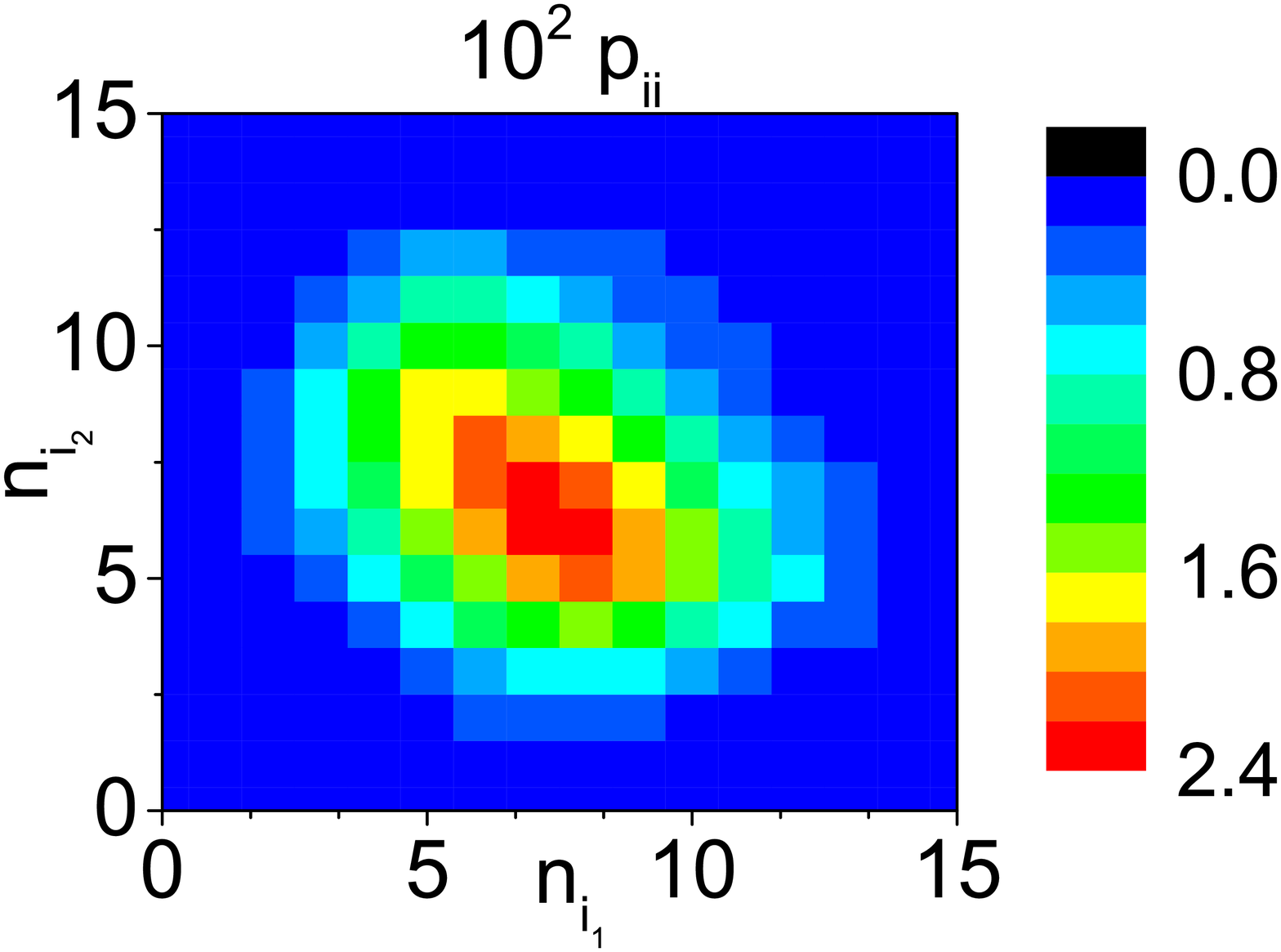}
 \includegraphics[width=0.48\hsize]{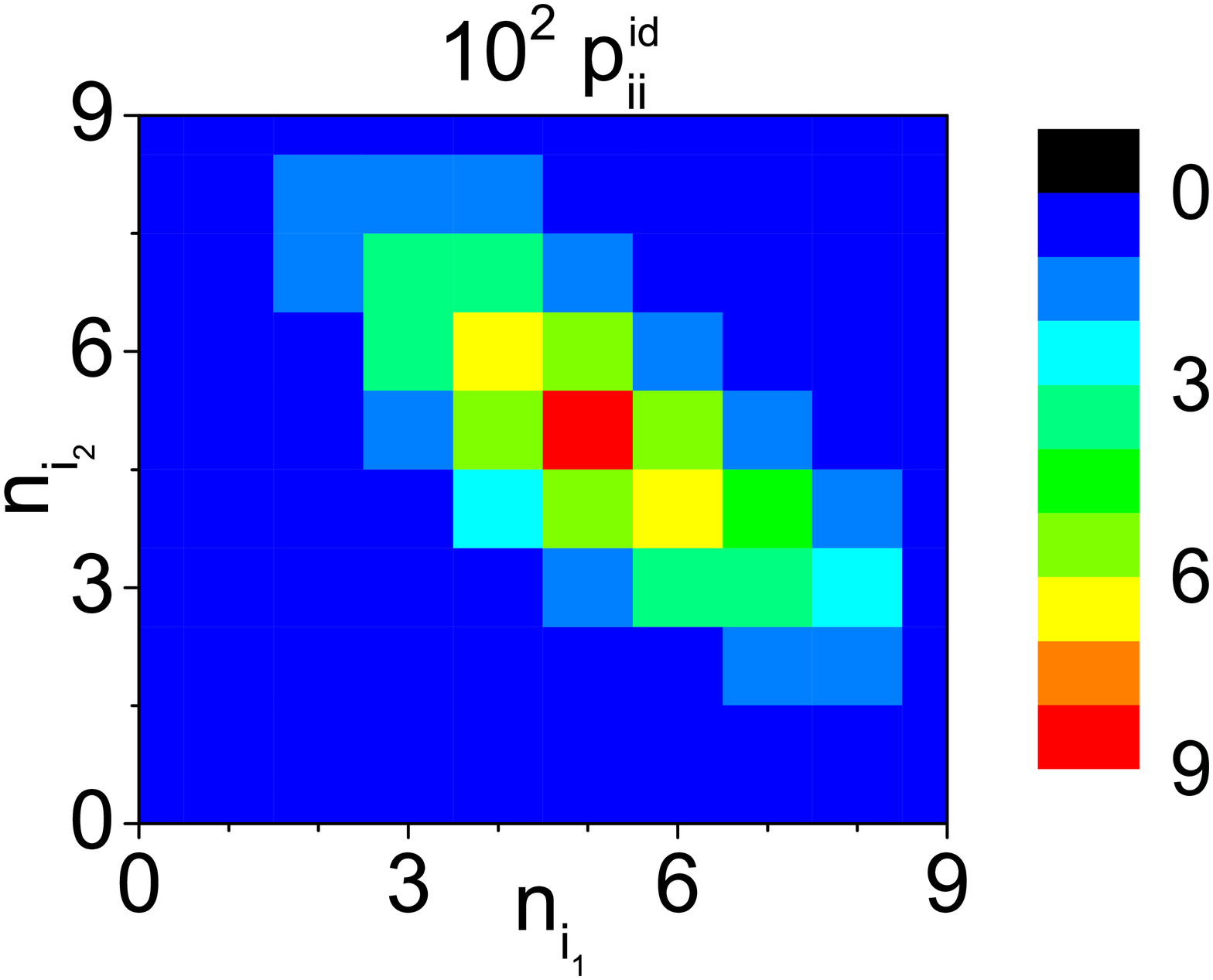}
  \vspace{2mm}
  \centerline{ \small (a) \hspace{.45\hsize} (b)}

 \includegraphics[width=0.48\hsize]{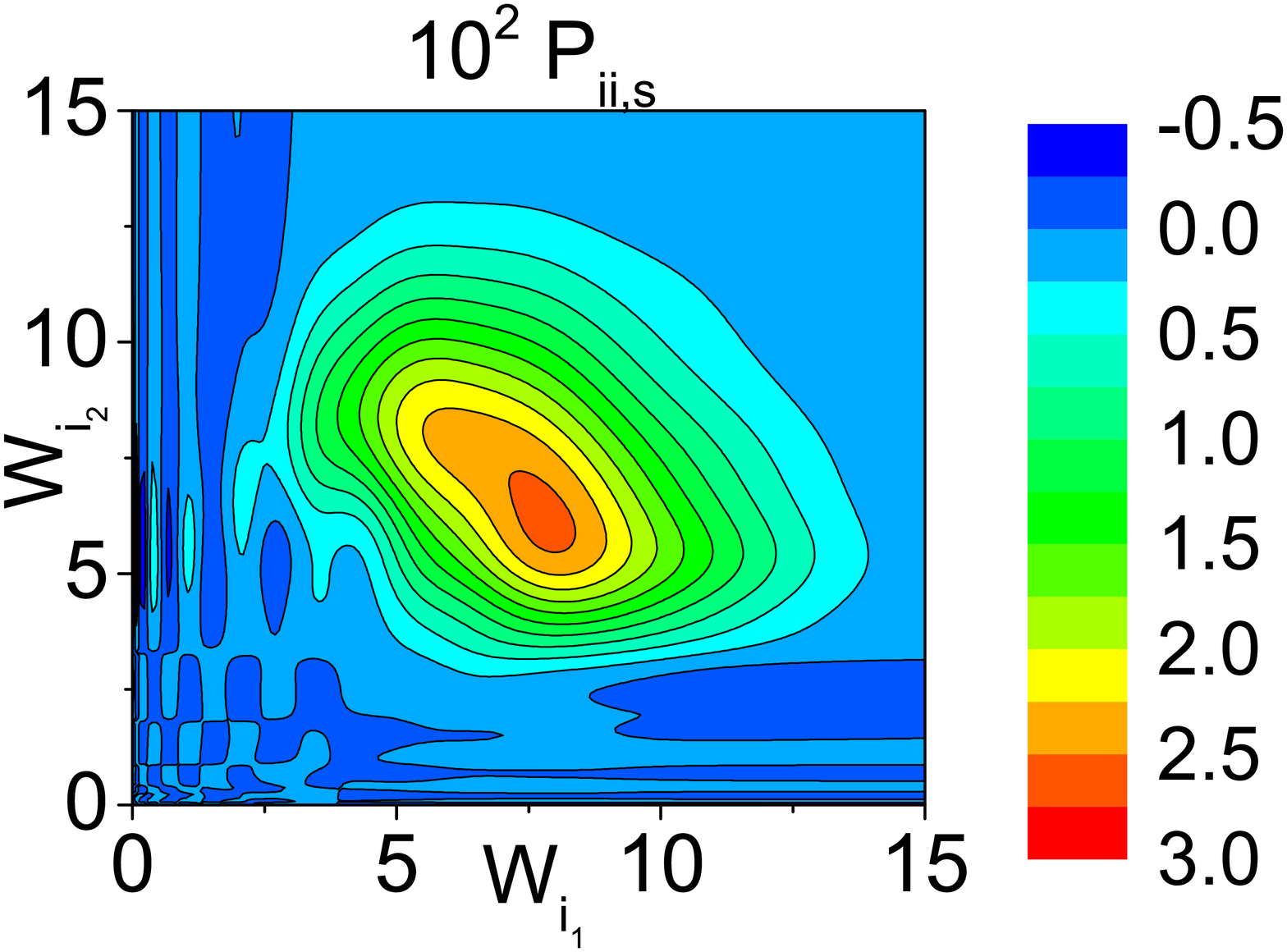}
 \includegraphics[width=0.48\hsize]{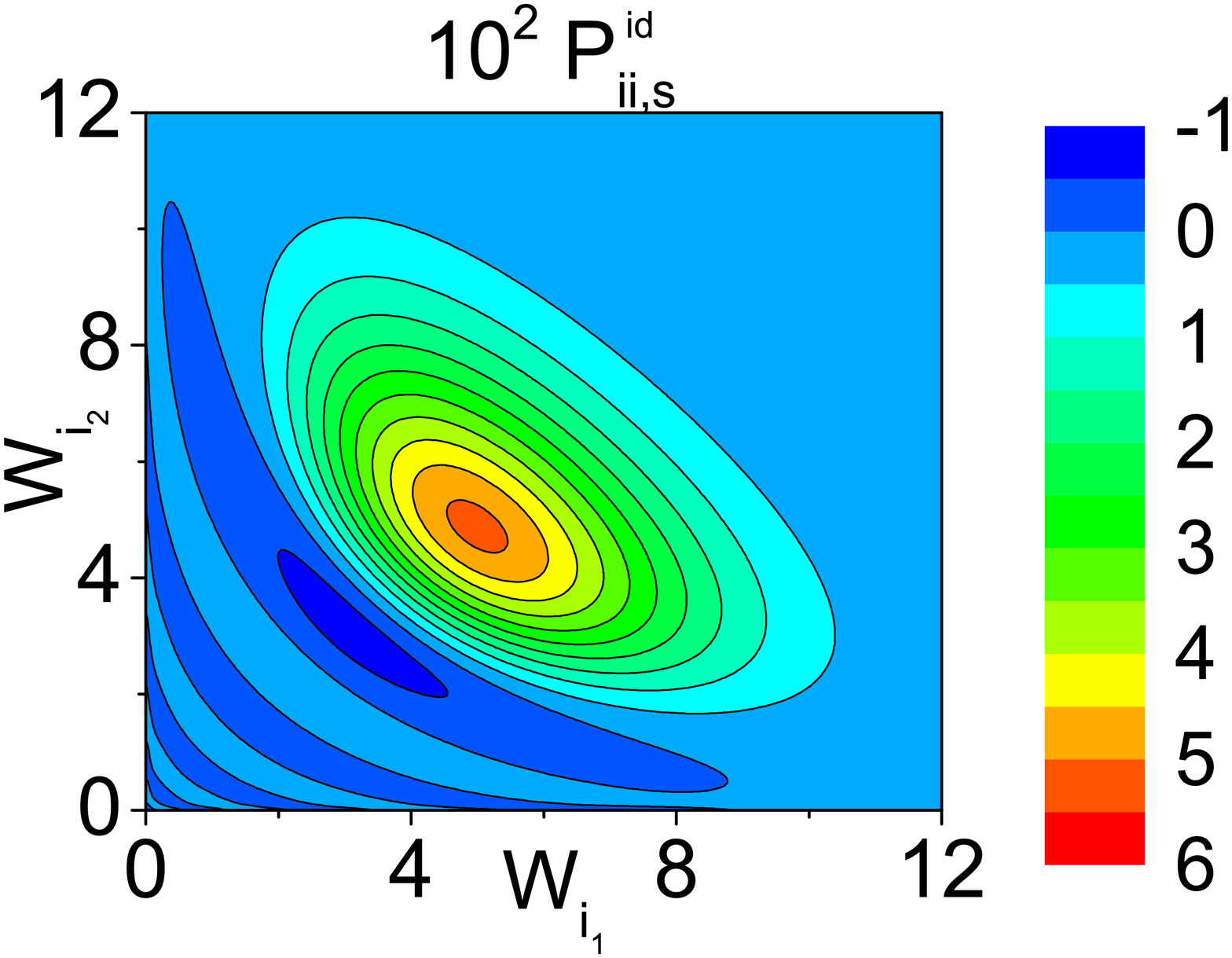}
  \vspace{2mm}
  \centerline{ \small (c) \hspace{.45\hsize} (d)}

 \includegraphics[width=0.48\hsize]{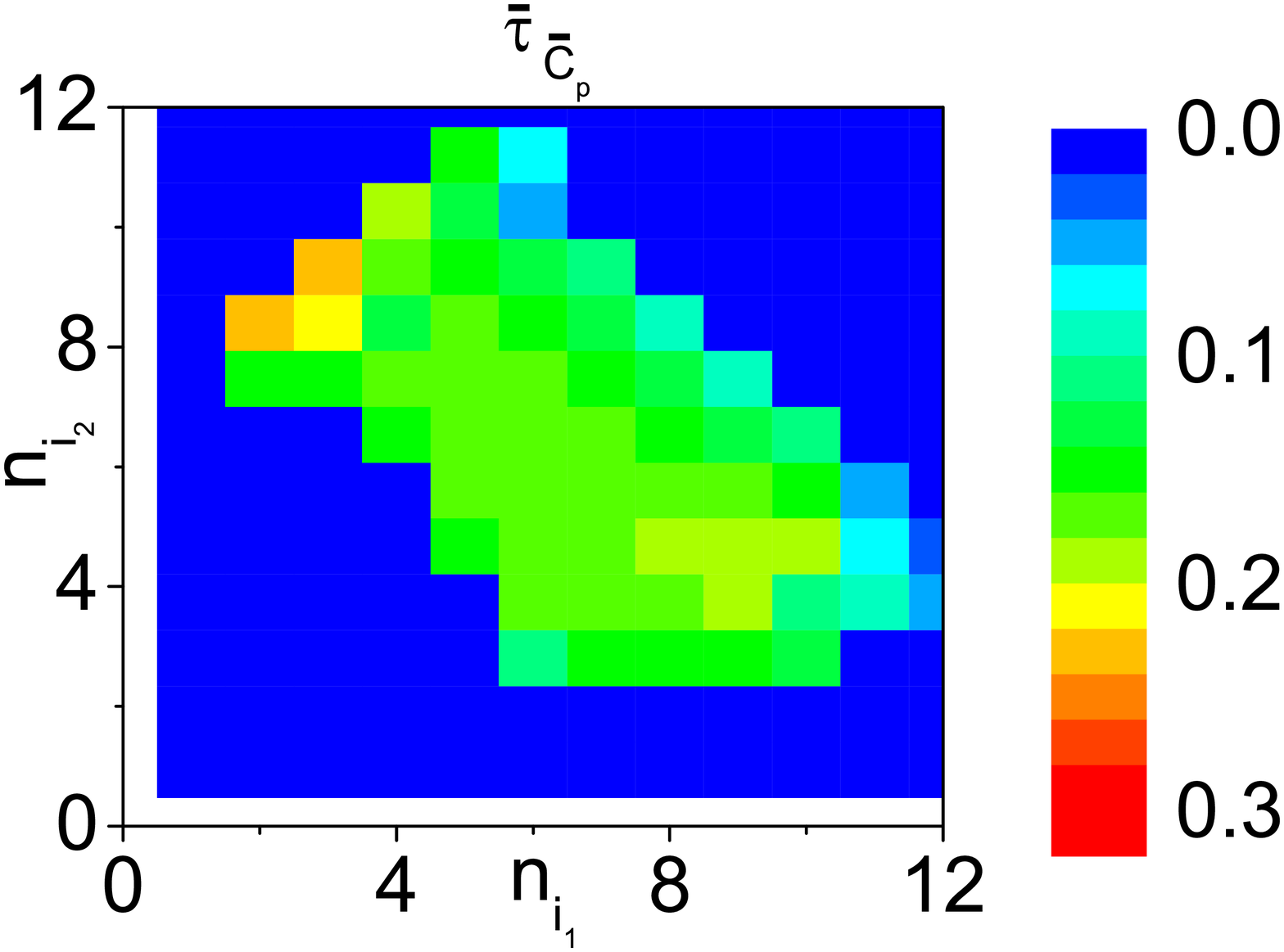}
 \includegraphics[width=0.48\hsize]{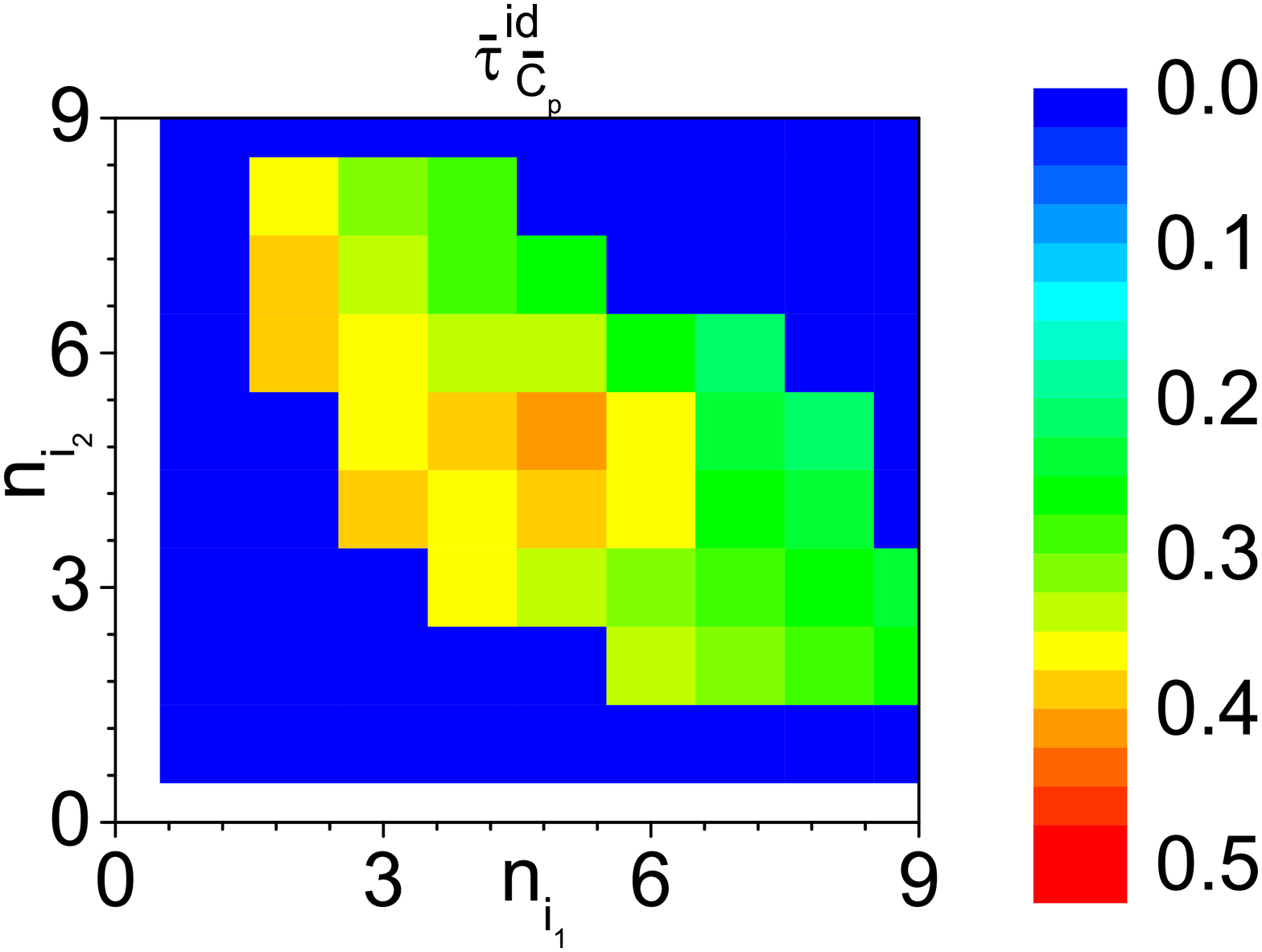}
  \vspace{2mm}
  \centerline{ \small (e) \hspace{.45\hsize} (f)}

 \includegraphics[width=0.48\hsize]{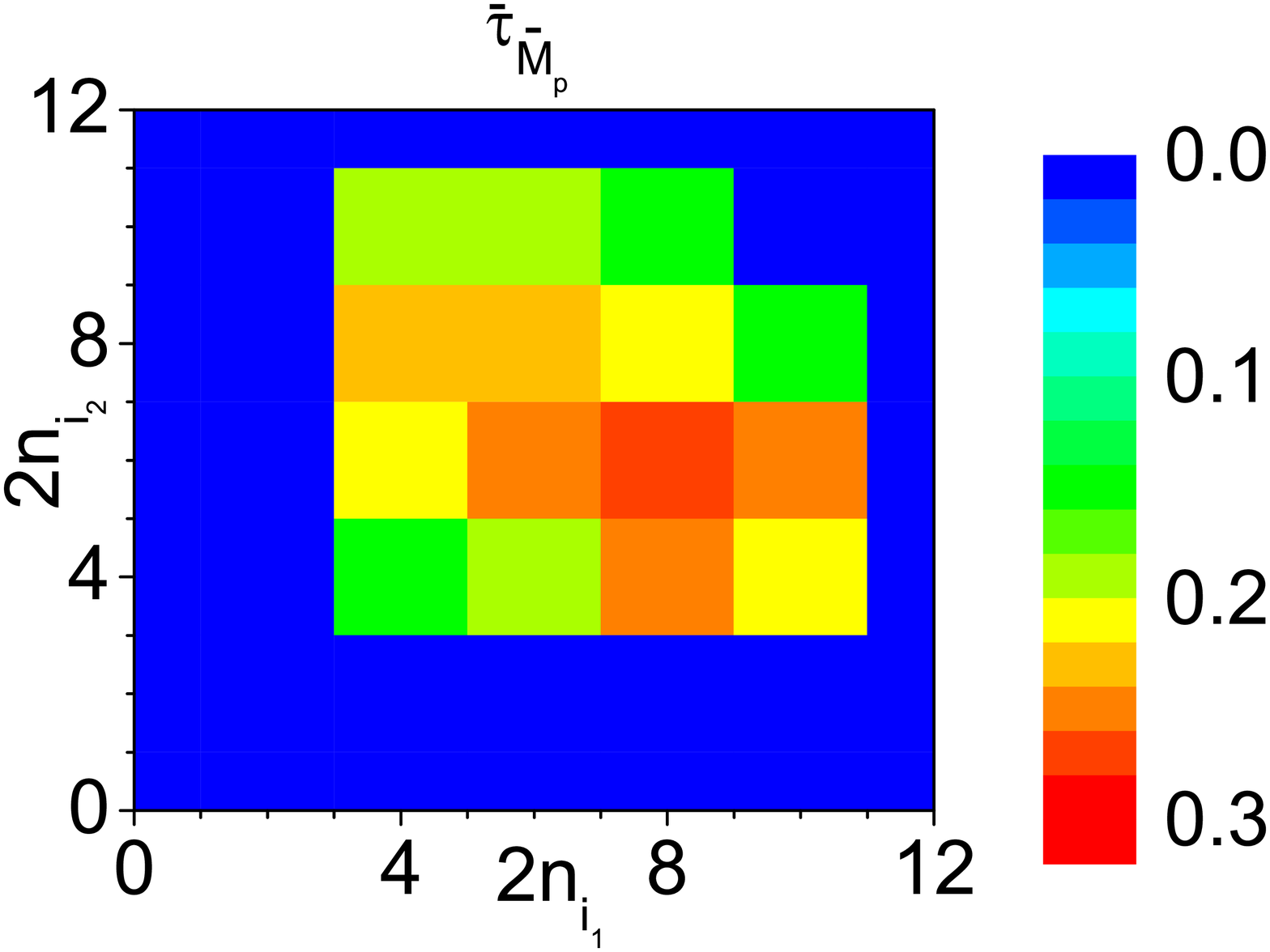}
 \includegraphics[width=0.48\hsize]{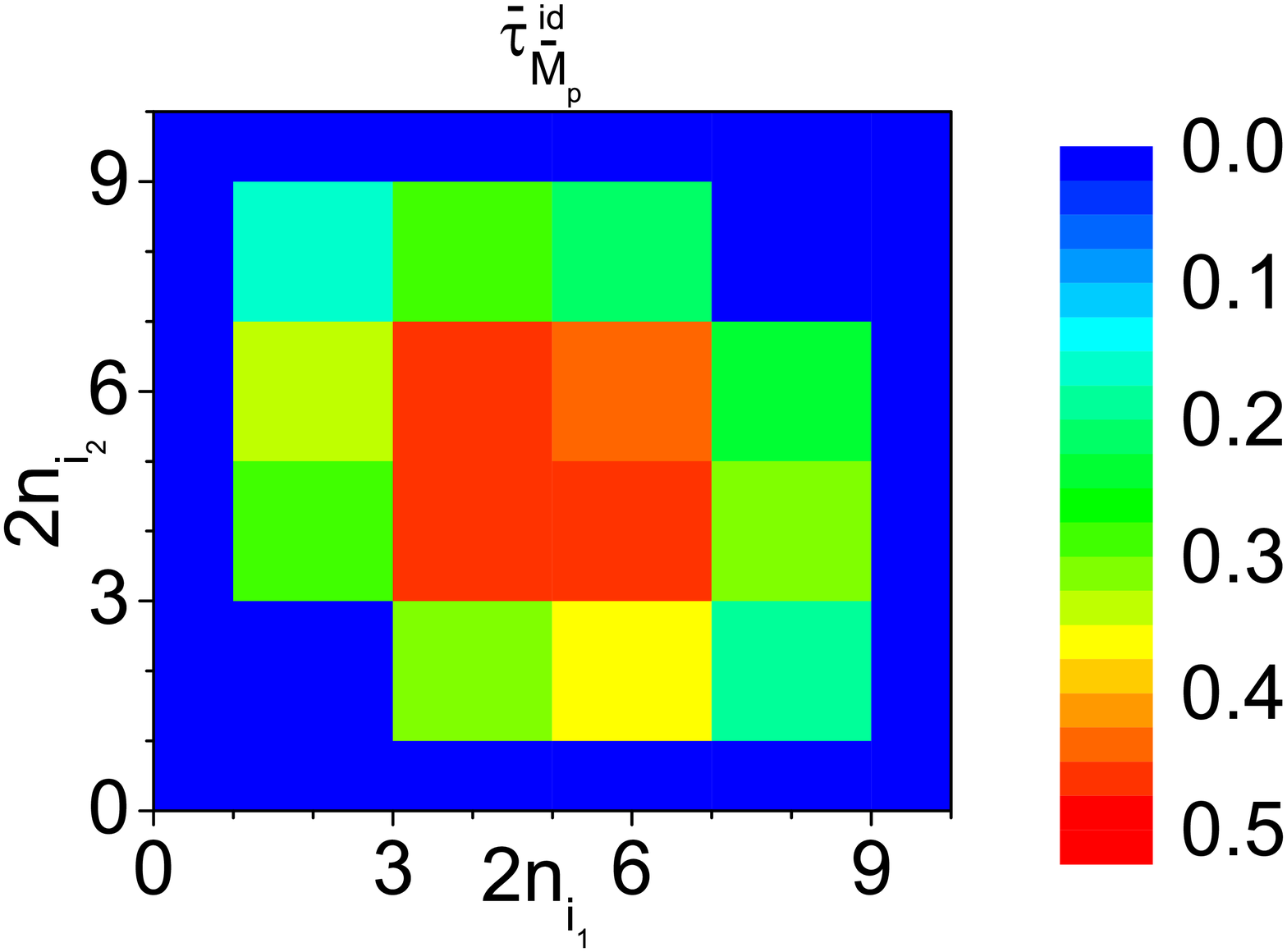}
  \vspace{2mm}
  \centerline{ \small (g) \hspace{.45\hsize} (h)}

 \includegraphics[width=0.48\hsize]{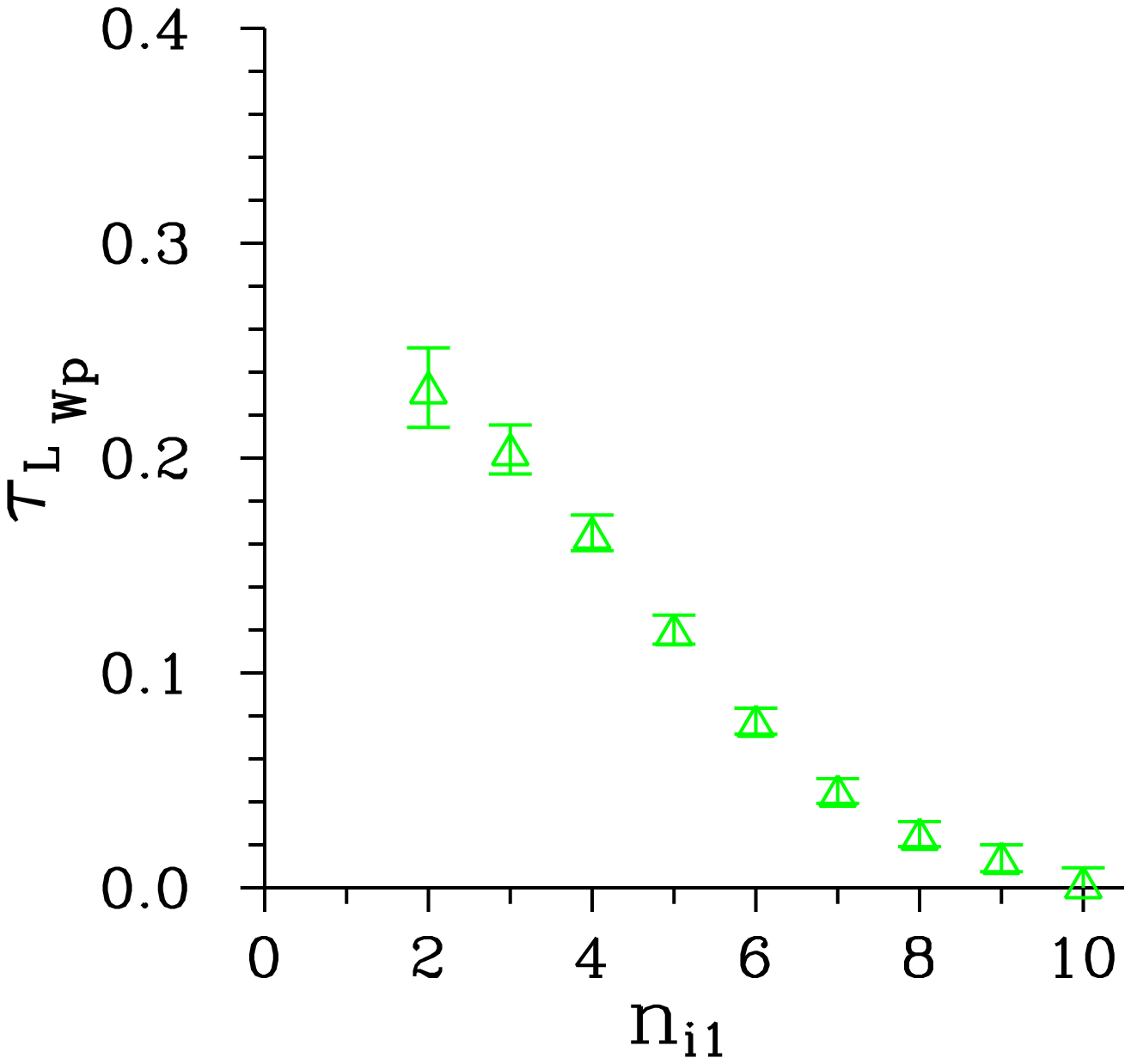}
 \includegraphics[width=0.48\hsize]{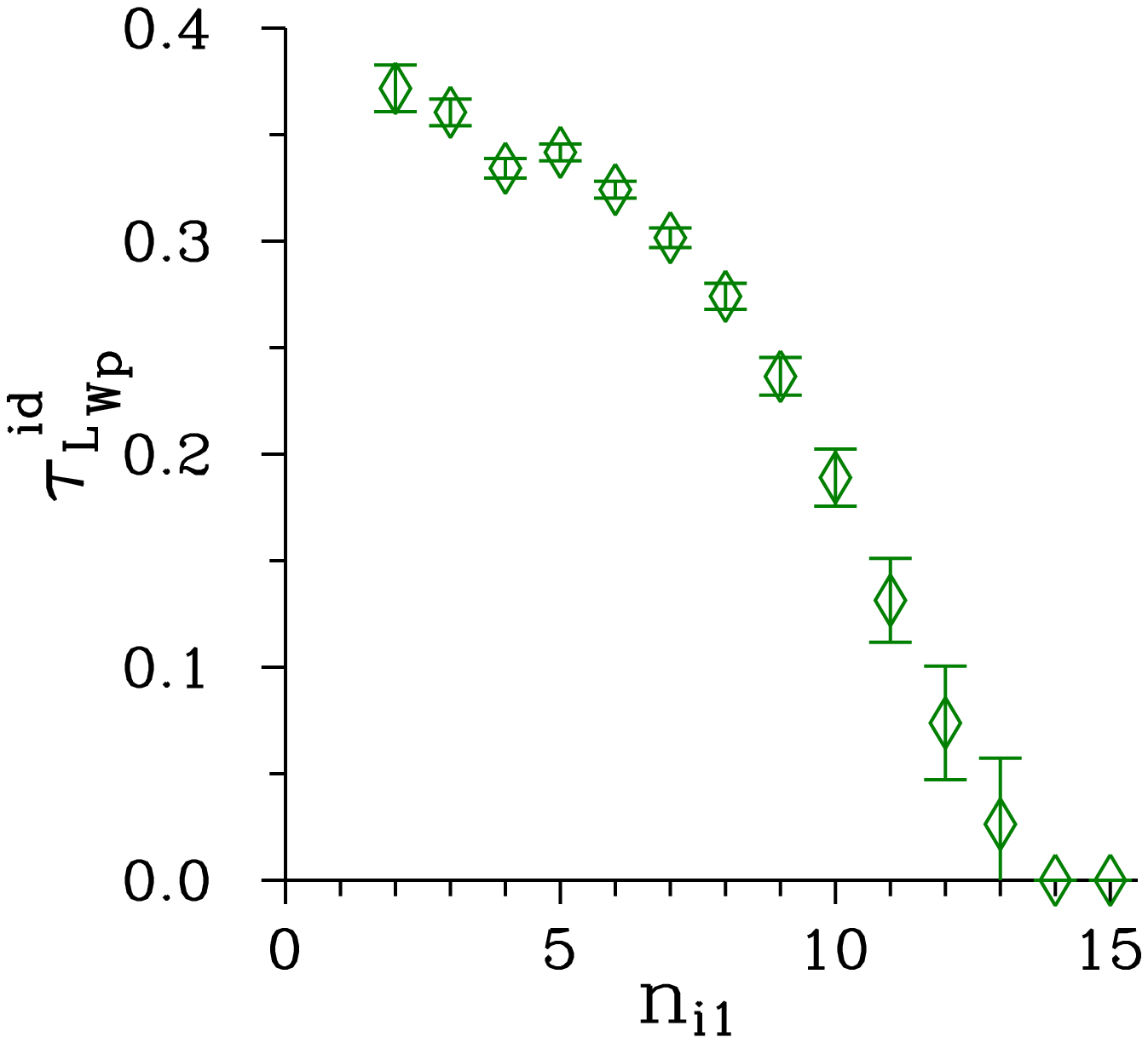}
  \vspace{2mm}
  \centerline{ \small (i) \hspace{.45\hsize} (j)}

 \caption{(a,b) Photon-number distribution $ p_{\rm ii}(n_{{\rm i}_1},n_{{\rm i}_2}) $ with (c,d) the corresponding
  quasi-distribution $ P_{{\rm ii},s}(W_{i_1},W_{i_2}) $ of integrated intensities and
  nonclassicality depths $ \bar{\tau} $ of the NCCa (e,f) $
  \bar{C}_{p}(n_{{\rm i}_1},n_{{\rm i}_2}) $, (g,h) $ \bar{M}_{p}(n_{{\rm i}_1},n_{{\rm i}_2}) $,
  and (i,j) $ L_{Wp}(n_{{\rm i}_1}) $ drawn as they depend on the numbers $ n_{{\rm i}_1} $
  and $ n_{{\rm i}_2} $ of photons in the idler fields. The fields postselected
  by $ c_{\rm s} = 5 $ signal photocounts (a,c,e,g,i) and $ n_{\rm s} = 10 $ signal photons
  (b,d,f,h,j) are analyzed. In (c) [(d)], $ s= 0.1 $ [$ s= -0.15 $] and the integrated intensities are expressed in the units of photon numbers.
  In (e,g) [(f,h)], only the
  NCCa for which the mean value of the used probabilities is greater than 0.01 [0.02] are considered.
  In (i) [(j)], isolated symbols (green $ \triangle $) [dark green $ \diamond $] originate in 2D [3D] maximum-likelihood
  method.}
\label{fig4}
\end{figure}

Now we compare side-by-side the properties of two typical postselected states
obtained by the real detector ($ c_{\rm s} = 5 $) and the ideal one ($ n_{\rm
s} = 10 $). The state generated in the experimental setup by the real detector
is a bit more intense, it contains on average around 7 photons in each idler
field compared to around 5 photons in the idler fields of the state provided by
the ideal detector. The correspoding 2D photon-number distributions $ p_{ii} $
and $ p_{ii}^{\rm id} $ plotted in Figs.~\ref{fig4}(a) and \ref{fig4}(b),
respectively, clearly exhibit prolongation in the direction perpendicular to
the line $ n_{{\rm i}_1} = n_{{\rm i}_2} $. Whereas the covariance $
C_{n,\Delta} $ of the idler-fields photon-number fluctuations $ \Delta n_{{\rm
i}_1} $ and $ \Delta n_{{\rm i}_2} $ equals only $ -0.14\pm 0.02 $ for the
state reached by the real detector, the ideal detector allows to reach the
value $-0.74\pm 0.03 $. Both these values belong to the nonclassical states as
the corresponding values of the modified noise-reduction-parameter are smaller
than 1 ($ R_{n,+} = 0.87\pm 0.03 $, $ R_{n,+}^{\rm id} = 0.16\pm 0.02 $). Also
the real detector provides the marginal idler fields with the classical
photon-number statistics close to the Poissonian one ($ F_{n,{\rm i}_1} = 1.03
\pm 0.09 $, $ F_{n,{\rm i}_2} = 1.01\pm 0.09 $). On the other hand, highly
sub-Poissonian states arise for the ideal detector ($ F_{n,{\rm i}_1}^{\rm id}
= 0.64\pm0.06 $, $ F_{n,{\rm i}_2}^{\rm id} = 0.61\pm 0.06 $). The NCCa $ C_W $
and $ M_W $ assign the NCDs $ \tau_W = 0.06\pm 0.02 $ ($ \tau_{C_W} = 0.02\pm
0.01 $, $ \tau_{M_W} = 0.06\pm 0.02 $) to the state obtained by the real
detector and $ 0.40 \pm 0.01 $ ($ \tau_{C_W}^{\rm id} = 0.38\pm 0.01 $, $
\tau_{M_W}^{\rm id} = 0.40 \pm 0.01 $) to the state provided by the ideal
detector.

The decomposition of quasi-distribution $ P_{{\rm ii},s}(W_{i_1},W_{i_2}) $ of
the idler-fields integrated intensities related to an arbitrary $ s $-ordering
of field operators into the Laguerre polynomials allows to reconstruct the
quasi-distribution $ P_{{\rm ii},s} $ from the corresponding photon-number
distribution $ p_{\rm ii}(n_{{\rm i}_1},n_{{\rm i}_2}) $ [for details, see
Appendix D]. The reconstructed quasi-distributions $ P_{\rm ii} $ and $ P_{\rm
ii}^{\rm id} $ belonging to the analyzed fields are drawn in
Figs.~\ref{fig4}(c) and \ref{fig4}(d) for $ s = 0.1 $ and $ s = -0.15 $,
respectively. As there occur negative values in both graphs and according to
the genuine definition of the nonclassicality \cite{Glauber1963,Sudarshan1963},
the actual NCDs $ \tau $ for the analyzed fields lie around 0.45 and 0.57 [see
Eq.~(\ref{10})], respectively. The areas with negative probability densities in
the plane $ (W_{i_1},W_{i_2}) $ are typically located in the region between the
point $ (W_{i_1},W_{i_2}) = (0,0) $ and the area where the maximal intensities
of the quasi-distribution $ P_{{\rm ii},s}(W_{i_1},W_{i_2}) $ occur [see the
graph in Fig.~\ref{fig4}(d)]. This resembles the behavior of 1D
quasi-distributions of integrated intensities characterizing sub-Poissonian
fields generated by photon-number-resolving postselection from TWBs
\cite{PerinaJr2013b}.

The values of the NCDs $ \tau $ indicated by negative values of the above
quasi-distributions are considerably greater than those revealed by the NCCa $
C_W $ and $ M_W $ based on the intensity moments, especially when the
photon-number distribution obtained by the real detector is analyzed. For this
reason, we extend our analysis of the nonclassicality by considering the
systems of NCCa involving the probabilities of photocount and photon-number
distributions. Also in this case, the systems of NCCa $ \bar{C}_p $ and $
\bar{M}_p $ derived from the Cauchy--Schwarz inequality and the matrix
approach, respectively, and described in detail in Appendix C proved the best
performance. Moreover, to certain extent, they revealed the location of
nonclassicality across the analyzed photon-number distributions, as
demonstrated in Figs.~\ref{fig4}(e--h) showing the corresponding NCDs $
\bar{\tau} $. The comparison of graphs in Figs.~\ref{fig4}(e,f) with those in
Figs.~\ref{fig4}(g,h) identifies the system of NCCa $ \bar{M}_p $ as more
powerful in quantifying the nonclassicality than the system of NCCa $ \bar{C}_p
$, similarly as in the case of their intensity-moment counterparts. The
attained values of the NCDs $ \bar{\tau}_{C_p} $ and $ \bar{\tau}_{M_p} $ are
greater than those reached by the NCCa $ C_W $ and $ M_W $ using the intensity
moments. Considerable improvement occurs for both photon-number distributions
($ \tau_{M_W} = 0.06\pm 0.02 $, $ \bar{\tau}_{M_p}^{\rm max} = 0.27 $; $
\tau_{M_W}^{\rm id} = 0.40\pm 0.01 $, $ \bar{\tau}_{M_p}^{\rm id,max} = 0.46
$). The greatest values of the NCDs $ \bar{\tau}_{M_p} $ are found in the
central parts of the photon-number distributions [see Figs.~\ref{fig4}(g,h)].

In our opinion, the NCCa based on the intensity moments lose their power to
resolve the nonclassicality compared to the NCCa with the probabilities in the
process of averaging that smoothes out the local nonclassical features
contained in the photon-number distributions. To support this explanation we
analyze both photon-number distributions applying the hybrid criterion $ L_{Wp}
$ in Eq.~(\ref{9}) that keeps the local 'resolution' in the first-idler-field
photon number $ n_{{\rm i}_1} $. The greatest achieved values of NCDs $
\tau_{L_{Wp}} $ and $ \tau_{L_{Wp}}^{\rm id} $ plotted in Figs.~\ref{fig4}(i)
and \ref{fig4}(j), respectively, are smaller than the corresponding greatest
values of the NCDs $ \bar{\tau}_{M_p}^{\rm max} $ and $ \bar{\tau}_{M_p}^{\rm
id,max} $ plotted in Figs.~\ref{fig4}(g) and \ref{fig4}(h), but they are
considerably greater than the values of the corresponding NCDs $ \tau_{M_W} $
and $ \tau_{M_W}^{\rm id} $.

\section{Conclusions}

Using postselection by a photon-number-resolving detector and two twin beams of
similar intensities, we have experimentally generated the fields with
increasing intensities that are endowed with anticorrelations in photon-number
fluctuations. They even exhibit the marginal sub-Poissonian photon-number
statistics under suitable conditions. Properties of the experimentally
generated postselected states were monitored by two additional
photon-number-resolving detectors. The obtained experimental data were
reconstructed in parallel by the maximum-likelihood approach and by considering
a suitable Gaussian fit. The nonclassicality of the observed postselected
fields was evidenced by the determination of the corresponding
quasi-distributions of integrated intensities with negative values as well as
by several types of the nonclassicality criteria and the accompanying
nonclassicality depths. Whereas the quasi-distributions of integrated
intensities are natural identifiers of the nonclassicality, the ability of the
nonclassicality criteria to resolve the nonclassicality decreases with their
decreasing 'resolution' (in turn, criteria based on the probabilities, hybrid
criteria and criteria using the intensity moments). Specific properties of the
generated states are appealing in quantum metrology: The measurement of
two-photon absorption cross-sections beyond the shot-noise-limit because of the
sub-Poissonian character of both fields and anticorrelations in photon-number
fluctuations serves as an example. The properties of the investigated states
are also attractive for two-photon excitations of molecules and other material
systems.

\acknowledgments The authors thank GA \v{C}R projects No. 18-08874S (V.M.,
R.M., O.H.) and No.~18-22102S (J.P.). They also acknowledge the support from
M\v{S}MT \v{C}R (project No. CZ.02.1.01/0.0/0.0/16\_019/0000754).

\appendix
\section{Multi-mode Gaussian fields and their reconstruction}

The mechanism of generation of the analyzed optical field suggests the
following analytical structure for its description. The optical field may be
considered as composed of two ideal multi-mode TWBs and three independent
multi-mode thermal (Gaussian) noisy fields. Spontaneous character of parametric
down-conversion suggests the photon-number distribution $ p_{{\rm p}_j}(n_{{\rm
s}_j},n_{{\rm i}_j}) $ for TWB $ j $, $ j=1,2 $, in the multi-mode Gaussian
form with $ M_{{\rm p}_j} $ modes and $ B_{{\rm p}_j} $ mean photon-pairs per
mode
\begin{equation}  
 p_{{\rm p}_j}(n_{{\rm s}_j},n_{{\rm i}_j}) = \delta_{n_{{\rm s}_j},n_{{\rm i}_j}}
  p^{\rm M-R}(n_{{\rm s}_j};M_{{\rm p}_j},B_{{\rm p}_j}).
\label{A1}
\end{equation}
The multi-mode thermal Mandel--Rice distribution $ p^{\rm M-R} $ for an $ M
$-mode field with each mode having on average $ B $ photons is given as:
\begin{equation}  
  p^{\rm M-R}(n;M,B) = \frac{ \Gamma(n+M) }{ n!\Gamma(M)} \frac{
   B^n}{(1+B)^{n+M}}.
\label{A2}
\end{equation}
In Eqs.~(\ref{A1}) and (\ref{A2}), the Kronecker symbol $ \delta_{n_{\rm
s},n_{i}} $ and the gamma function $ \Gamma $ are used.

A 3D photon-number distribution $ p_{\rm p} $ of the ideally paired part of the
studied optical field is expresses as:
\begin{eqnarray}  
 p_{\rm p}(n_{\rm s},n_{{\rm i}_1},n_{{\rm i}_2}) = \sum_{n_{{\rm s}_1}=0}^{n_{\rm s}} p_{{\rm p}_1}(n_{{\rm s}_1},n_{{\rm i}_1})
  p_{{\rm p}_2}(n_{\rm s}-n_{{\rm s}_1},n_{{\rm i}_2}). \nonumber \\
 & &
\label{A3}
\end{eqnarray}
We assume the photon-number distribution $ p_{\rm n_{\rm s}} $ of the noise in
the combined signal field in the form of Eq.~(\ref{A2}) with $ M_{\rm n_{\rm
s}} $ modes each having on average $ B_{\rm n_{\rm s}} $ noisy photons. Similar
assumption is made for the photon-number distribution $ p_{\rm n_{{\rm i}_j}} $
of the $ j $-th idler field whose noise is distributed into $ M_{\rm n_{{\rm
i}_j}} $ modes each populated with $ B_{\rm n_{{\rm i}_j}} $ mean photons, $
j=1,2 $. Three-fold convolution of the ideally paired photon-number
distribution $ p_{\rm p} $ with three noisy photon-number distributions then
leaves us with the photon-number distribution $ p $ appropriate for the
analyzed optical field:
\begin{eqnarray} 
 p(n_{\rm s},n_{{\rm i}_1},n_{{\rm i}_2}) &=& \sum_{l_{\rm s}=0}^{n_{\rm s}} p_{\rm n_{\rm s}}(n_{\rm s}-l_{\rm s})
  \sum_{l_{{\rm i}_1}=0}^{n_{{\rm i}_1}} p_{\rm n_{{\rm i}_1}}(n_{{\rm i}_1}-l_{{\rm i}_1})
  \nonumber \\
 & & \hspace{-5mm}  \times  \sum_{l_{{\rm i}_2}=0}^{n_{{\rm i}_2}} p_{\rm n_{{\rm i}_2}}(n_{{\rm i}_2}-l_{{\rm i}_2})
   p_{\rm p}(l_{\rm s},l_{{\rm i}_1},l_{{\rm i}_2}).
\label{A4}
\end{eqnarray}

The photon-number moments $ \langle n_{\rm s}^{k_{\rm s}} n_{{\rm
i}_1}^{k_{{\rm i}_1}} n_{{\rm i}_2}^{k_{{\rm i}_2}} \rangle $ corresponding to
the photon-number distribution $ p $ in Eq.~(\ref{A4}) are determined as
follows:
\begin{eqnarray}   
 \langle n_{\rm s}^{k_{\rm s}} n_{{\rm i}_1}^{k_{{\rm i}_1}} n_{{\rm i}_2}^{k_{{\rm i}_2}} \rangle &=&
  \sum_{n_{\rm s},n_{{\rm i}_1},n_{{\rm i}_2}=0}^{\infty}
  n_{\rm s}^{k_{\rm s}} n_{{\rm i}_1}^{k_{{\rm i}_1}} n_{{\rm i}_2}^{k_{{\rm i}_2}}
  p(n_{\rm s},n_{{\rm i}_1},n_{{\rm i}_2}). \nonumber \\
 & &
\label{A5}
\end{eqnarray}
The (integrated-) intensity moments $ \langle W_{\rm s}^{k_{\rm s}} W_{{\rm
i}_1}^{k_{{\rm i}_1}} W_{{\rm i}_2}^{k_{{\rm i}_2}} \rangle $, that are the
normally-ordered photon-number moments, are derived from the above
photon-number moments using the Stirling numbers $ S $ of the first kind
\cite{Gradshtein2000}:
\begin{eqnarray}   
 \langle W_{\rm s}^{k_{\rm s}} W_{{\rm i}_1}^{k_{{\rm i}_1}} W_{{\rm i}_2}^{k_{{\rm i}_2}} \rangle &=&
  \sum_{l_{\rm s}=0}^{k_{\rm s}} S(k_{\rm s},l_{\rm s})
  \sum_{l_{{\rm i}_1}=0}^{k_{{\rm i}_1}} S(k_{{\rm i}_1},l_{{\rm i}_1}) \nonumber \\
 & & \hspace{-5mm} \times \sum_{l_{{\rm i}_2}=0}^{k_{{\rm i}_2}} S(k_{{\rm i}_2},l_{{\rm i}_2})
   \langle n_{\rm s}^{l_{\rm s}} n_{{\rm i}_1}^{l_{{\rm i}_1}} n_{{\rm i}_2}^{l_{{\rm i}_2}} \rangle.
\label{A6}
\end{eqnarray}
The inverse relation to that in Eq.~(\ref{A6}) relies on the Stirling numbers
of the second kind. We note that we have the following relations between the
intensity moments and number $ M $ of modes together with their mean photon
numbers $ B $ for a multi-mode thermal field:
\begin{eqnarray}  
 B = \frac{\langle (\Delta W)^2 \rangle }{ \langle W\rangle } , \hspace{4mm}
 M = \frac{ \langle W \rangle^2 }{ \langle (\Delta W)^2 \rangle};
\label{A7}
\end{eqnarray}
$ \Delta W \equiv W - \langle W\rangle $.

In the experiment, we detect the photocount numbers $ c $, i.e. the numbers of
photoelectons excited by the absorbed photons. Multiple realizations of the
measurement then give us the experimental photocount histogram $ f $ determined
in Eq.~(\ref{3}) and the accompanying photocount moments $ \langle c_{\rm
s}^{k_{\rm s}} c_{{\rm i}_1}^{k_{{\rm i}_1}} c_{{\rm i}_2}^{k_{{\rm i}_2}}
\rangle $,
\begin{eqnarray}   
 \langle c_{\rm s}^{k_{\rm s}} c_{{\rm i}_1}^{k_{{\rm i}_1}} c_{{\rm i}_2}^{k_{{\rm i}_2}} \rangle =
  \sum_{c_{\rm s},c_{{\rm i}_1},c_{{\rm i}_2}=0}^{\infty}
  c_{\rm s}^{k_{\rm s}} c_{{\rm i}_1}^{k_{{\rm i}_1}} c_{{\rm i}_2}^{k_{{\rm i}_2}}
  f(c_{\rm s},c_{{\rm i}_1},c_{{\rm i}_2}).
\label{A8}
\end{eqnarray}
Similarly as the intensity moments $ \langle W_{\rm s}^{k_{\rm s}} W_{{\rm
i}_1}^{k_{{\rm i}_1}} W_{{\rm i}_2}^{k_{{\rm i}_2}} \rangle $ are assigned to
the photon-number moments $ \langle n_{\rm s}^{l_{s}} n_{{\rm i}_1}^{l_{{\rm
i}_1}} n_{{\rm i}_2}^{l_{{\rm i}_2}} \rangle $, we may assign the intensity
moments $ \langle {\cal W}_{\rm s}^{k_{\rm s}} {\cal W}_{{\rm i}_1}^{k_{{\rm
i}_1}} {\cal W}_{{\rm i}_2}^{k_{{\rm i}_2}} \rangle_E $ to the photocount
moments $ \langle c_{\rm s}^{l_{\rm s}} c_{{\rm i}_1}^{l_{{\rm i}_1}} c_{{\rm
i}_2}^{l_{{\rm i}_2}} \rangle $ using the relations in Eq.~(\ref{A6}). The
photocount moments $ \langle c_{\rm s}^{l_{\rm s}} c_{{\rm i}_1}^{l_{{\rm
i}_1}} c_{{\rm i}_2}^{l_{{\rm i}_2}} \rangle $ as well as the intensity moments
$ \langle {\cal W}_{\rm s}^{k_{\rm s}} {\cal W}_{{\rm i}_1}^{k_{{\rm i}_1}}
{\cal W}_{{\rm i}_2}^{k_{{\rm i}_2}} \rangle_E $ are directly available from
the experimental data and so they form a natural basis for the reconstruction
of the above Gaussian form of the studied field.

Description of the response of a PNRD is also needed when making the
reconstruction. An iCCD camera, used in our experiment, is characterized by
detection efficiency $ \eta $, dark-count rate $ D \equiv d / N $ per pixel and
number $ N $ of active pixels that determine the corresponding detection matrix
$ T(c,n) $ introduced in Eq.~(\ref{2}) in the following form
\cite{PerinaJr2012}:
\begin{eqnarray}     
 T(c,n) &=& \left( \begin{array}{c} N \cr c \end{array} \right)
  (1-D)^{N} (1-\eta)^{n} (-1)^{c} \nonumber \\
 & & \hspace{-10mm} \times
  \sum_{l=0}^{c}
  \left( \begin{array}{c} c \cr l \end{array} \right) \frac{(-1)^l}{(1-D)^l}
  \left( 1 + \frac{l}{N} \frac{\eta}{1-\eta}
   \right)^{n} .
\label{A9}
\end{eqnarray}

For the reconstruction, we have at our disposal the experimental 3D photocount
histogram $ f $. From this histogram, we conveniently determine the following
nine experimental intensity moments with sufficiently high precision: $ \langle
{\cal W}_{\rm s}\rangle_E $, $ \langle{\cal W}_{{\rm i}_j}\rangle_E $, $
\langle(\Delta {\cal W}_{\rm s})^2\rangle_E $, $ \langle(\Delta {\cal W}_{{\rm
i}_j})^2\rangle_E $, $ \langle \Delta {\cal W}_{\rm s} \Delta {\cal W}_{{\rm
i}_j} \rangle_E $, and $ \langle \Delta {\cal W}_{{\rm i}_1} \Delta {\cal
W}_{{\rm i}_2} \rangle_E $, $ j=1,2 $. On the other hand, the multi-mode
Gaussian optical field is characterized by ten parameters, five parameters give
the numbers of modes ($ M_{{\rm p}_j} $, $ M_{\rm n_{\rm s}} $, $ M_{{\rm
n}_{{\rm i}_j}} $, $ j=1,2 $) in different components of the field and five
parameters characterize the mean photon (-pair) numbers in each mode ($ B_{{\rm
p}_j} $, $ B_{\rm n_{\rm s}} $, $ B_{{\rm n}_{{\rm i}_j}} $, $ j=1,2 $).
Moreover, we need to know the detection efficiencies for each detected field ($
\eta_{\rm s} $, $ \eta_{{\rm i}_1} $, $ \eta_{{\rm i}_2} $).

Detailed analysis of the used experimental setup reveals that the detection
efficiencies $ \eta_{{\rm i}_1} $ and $ \eta_{{\rm i}_2} $ cannot be determined
independently with sufficient precision. This is related to the fact that no
photon pairs occur directly in the first and the second idler fields. For this
reason, we assume in our analysis that they equal ($ \eta_{{\rm i}_1} =
\eta_{{\rm i}_2} \equiv \eta_{\rm i} $). Under this assumption we can
accomplish the reconstruction in two subsequent steps.

First, we combine together the intensity moments of both idler fields to arrive
at the moments characterizing the common idler field:
\begin{eqnarray}   
 \langle {\cal W}_{\rm i}\rangle_E &=& \langle {\cal W}_{{\rm i}_1}\rangle_E
  + \langle {\cal W}_{{\rm i}_2}\rangle_E, \nonumber \\
 \langle (\Delta {\cal W}_{\rm i})^2\rangle_E &=& \langle (\Delta {\cal W}_{{\rm i}_1})^2\rangle_E
  + 2\langle \Delta{\cal W}_{{\rm i}_1}\Delta{\cal W}_{{\rm i}_2}\rangle_E \nonumber \\
 & & + \langle(\Delta {\cal W}_{{\rm i}_2})^2\rangle_E, \nonumber \\
 \langle \Delta {\cal W}_{\rm s}\Delta {\cal W}_{\rm i}\rangle_E &=&
  \langle \Delta {\cal W}_{\rm s}\Delta {\cal W}_{{\rm i}_1}\rangle_E +
  \langle \Delta {\cal W}_{\rm s}\Delta {\cal W}_{{\rm i}_2}\rangle_E.  \nonumber \\
 & &
\label{A10}
\end{eqnarray}
Then we apply the reconstruction method for a multi-mode Gaussian TWB composed
of the combined signal and combined idler fields that has been developed
in \cite{PerinaJr2013a}. This provides us the intensity moments $ \langle W_{\rm
p}\rangle $ and $ \langle (\Delta W_{\rm p})^2\rangle $ of the combined ideally
paired field and intensity moments $ \langle W_{\rm n_{\rm s}}\rangle $, $
\langle W_{{\rm n}_{\rm i}}\rangle $, $ \langle (\Delta W_{\rm n_{\rm
s}})^2\rangle $, and $ \langle (\Delta W_{{\rm n}_{\rm i}})^2\rangle $ of the
noise signal and idler fields as well as the detection efficiencies $ \eta_{\rm
s} $ and $ \eta_{\rm i} $.

In the second step, we determine the remaining intensity moments $ \langle
W_{{\rm p}_j}\rangle $ and $ \langle (\Delta W_{{\rm p}_j})^2\rangle $
belonging to the paired components as well as the intensity moments $ \langle
W_{{\rm i}_j}\rangle $ and $ \langle (\Delta W_{{\rm i}_j})^2\rangle $ of the
noise idler fields, $ j=1,2 $. For this purpose, we write the following ten
linear relations among the looked-for intensity moments:
\begin{eqnarray}   
 \langle W_{{\rm p}_j}\rangle + \langle W_{{\rm n}_{{\rm i}_j}}\rangle &=& \langle {\cal W}_{{\rm i}_j}\rangle_E
  / \eta_{\rm i} , \nonumber \\
 \langle (\Delta W_{{\rm p}_j})^2\rangle + \langle (\Delta W_{{\rm i}_j})^2\rangle &=& \langle (\Delta {\cal W}_{{\rm i}_j})^2 \rangle_E
  / \eta_{\rm i}^2 , \nonumber \\
 \langle W_{{\rm p}_j}\rangle + \langle (\Delta W_{{\rm p}_j})^2\rangle &=& \langle \Delta {\cal W}_{{\rm i}_j}\Delta {\cal W}_{\rm s} \rangle_E
  / (\eta_{\rm i} \eta_{\rm s}) , \nonumber \\
  & & \hspace{20mm} j=1,2, \nonumber \\
 \langle W_{{\rm p}_1}\rangle + \langle W_{{\rm p}_2}\rangle &=& \langle W_{\rm p}\rangle,
   \nonumber \\
 \langle W_{{\rm n}_{{\rm i}_1}}\rangle + \langle W_{{\rm n}_{{\rm i}_2}}\rangle &=& \langle W_{{\rm n}_{\rm i}}\rangle,
   \nonumber \\
 \langle (\Delta W_{{\rm p}_1})^2\rangle + \langle (\Delta W_{{\rm p}_2})^2\rangle &=& \langle (\Delta W_{\rm p})^2\rangle,
  \nonumber \\
 \langle (\Delta W_{{\rm n}_{{\rm i}_1}})^2\rangle + \langle (\Delta W_{{\rm n}_{{\rm i}_2}})^2\rangle &=& \langle (\Delta
 W_{{\rm n}_{\rm i}})^2\rangle.
 \label{A11}
\end{eqnarray}
Whereas the first six relations in Eq.~(\ref{A11}) contain the original
experimental intensity moments, the remaining four relations are based upon the
intensity moments obtained in the first step.

Detailed analysis of the linear relations in Eq.~(\ref{A11}) reveals that only
seven out of them are independent. As we have eight independent intensity
moments to be determined, we choose one intensity moment as a free parameter
and derive the remaining seven ones using the relations in Eq.~(\ref{A11}). We
may conveniently choose, e.g., the moment $ \langle (\Delta W_{{\rm
p}_1})^2\rangle $ and express the remaining moments as linear combinations of
this moment, the experimental intensity moments and the moments known from the
first step. We may proceed, e.g., along the following lines: $ \langle (\Delta
W_{{\rm p}_1})^2\rangle \rightarrow \langle W_{{\rm p}_1}\rangle \rightarrow
\langle W_{{\rm p}_2}\rangle \rightarrow \langle (\Delta W_{{\rm
p}_2})^2\rangle $, $ \langle W_{{\rm p}_j}\rangle \rightarrow \langle W_{{\rm
n}_{{\rm i}_j}}\rangle $, $ \langle (\Delta W_{{\rm p}_j})^2\rangle \rightarrow
\langle (\Delta W_{{\rm n}_{{\rm i}_j}})^2\rangle $, $ j=1,2 $. We note that
the allowed values of the intensity moment $ \langle (\Delta W_{{\rm
p}_1})^2\rangle $ fulfill:
\begin{eqnarray}   
 \langle (\Delta W_{{\rm p}_1})^2\rangle &\in & (0, \min\{ \langle (\Delta {\cal W}_{{\rm i}_1})^2 \rangle_E
  / \eta_{\rm i}^2, \nonumber \\
 & & \hspace{5mm}  \langle \Delta {\cal W}_{{\rm i}_1}\Delta {\cal W}_{\rm s} \rangle_E
  / (\eta_{\rm i} \eta_{\rm s}) \} ) .
\label{A12}
\end{eqnarray}
For given set of the values of the intensity moments $ \langle W_{{\rm p}_j}
\rangle $, $ \langle (\Delta W_{{\rm p}_j})^2 \rangle $, $ \langle W_{{\rm
n}_{{\rm i}_j}} \rangle $, $ \langle (\Delta W_{{\rm n}_{{\rm i}_j}})^2 \rangle
$, $ j=1,2 $, $ \langle W_{n_{\rm s}} \rangle $, and $ \langle (\Delta
W_{n_{\rm s}})^2 \rangle $ we derive the numbers $ M_{{\rm p}_j} $, $ M_{{\rm
n}_{{\rm i}_j}} $, $ j=1,2 $, and $ M_{n_{\rm s}} $ of modes and mean photon
(-pair) numbers $ B_{{\rm p}_j} $, $ B_{{\rm n}_{{\rm i}_j}} $, $ j=1,2 $, and
$ B_{n_{\rm s}} $ using Eqs.~(\ref{A7}). Then, we reconstruct the 3D photon
number distribution $ p(n_{\rm s},n_{{\rm i}_1},n_{{\rm i}_2}) $ in
Eq.~(\ref{A3}) and arrive at the theoretical 3D photocount histogram $ f^{\rm
th}(c_{\rm s},c_{{\rm i}_1},c_{{\rm i}_2}) $ by applying Eq.~(\ref{3}) together
with the detection matrix in Eq.~(\ref{A9}). The optimal values of numbers of
modes and mean photon (-pair) numbers are set such that they minimize the
declination function $ {\cal D} $ between the theoretical and experimental
histograms:
\begin{equation}   
 {\cal D} = \sqrt{ \sum_{c_{\rm s},c_{{\rm i}_1},c_{{\rm i}_2}=0}^{\infty}[ f^{\rm th}(c_{\rm s},c_{{\rm i}_1},c_{{\rm i}_2}) -
 f(c_{\rm s},c_{{\rm i}_1},c_{{\rm i}_2})]^2 }.
\label{A13}
\end{equation}

\section{Maximum-likelihood reconstruction of 2D and 3D photon-number distributions}

The 3D photon-number distribution $ p(n_{\rm s},n_{{\rm i}_1},n_{{\rm i}_2}) $
of the original optical field used in the experiment is obtained from the
experimental photocount histogram $ f(c_{\rm s},c_{{\rm i}_1},c_{{\rm i}_2}) $
by inverting the linear relations expressed in Eq.~(\ref{3}). The
maximum-likelihood method \cite{Dempster1977,Vardi1993} provides us the
following iteration procedure that reveals the photon-number distribution $
p(n_{\rm s},n_{{\rm i}_1},n_{{\rm i}_2}) $ as a steady state of the following
iteration procedure:
\begin{eqnarray}   
 p^{(j+1)}(n_{\rm s},n_{{\rm i}_1},n_{{\rm i}_2})&=& \sum_{c_{\rm s},c_{{\rm i}_1},c_{{\rm i}_2}=0}^{\infty}
  F^{(j)}(c_{\rm s},c_{{\rm i}_1},c_{{\rm i}_2})  T_{\rm s}(c_{\rm s},n_{\rm s}) \nonumber \\
 & & \times T_{{\rm i}_1}(c_{{\rm i}_1},n_{{\rm i}_1}) T_{{\rm i}_2}(c_{{\rm i}_2},n_{{\rm i}_2}),
\label{B1} \\
 F^{(j)}(c_{\rm s},c_{{\rm i}_1},c_{{\rm i}_2}) &=& f(c_{\rm s},c_{{\rm i}_1},c_{{\rm i}_2}) \Biggl[
  \sum_{n'_{\rm s},n'_{{\rm i}_1},n'_{{\rm i}_2}=0}^{\infty} T_{\rm s}(c_{\rm s},n'_{\rm s}) \nonumber \\
 & & \hspace{-18mm} \times  T_{{\rm i}_1}(c_{{\rm i}_1},n'_{{\rm i}_1})
  T_{{\rm i}_2}(c_{{\rm i}_2},n'_{{\rm i}_2}) p^{(j)}(n'_{\rm s},n'_{{\rm i}_1},n'_{{\rm i}_2}) \Bigr]^{-1}, \nonumber \\
 & &   \hspace{2mm} j=0,1, \ldots \; .
 \nonumber
\end{eqnarray}

Similarly, the 2D photon-number distributions $ p_{\rm ii}(n_{{\rm
i}_1},n_{{\rm i}_2};c_{\rm s}) $ given in Eq.~(\ref{2}) and belonging to the
field postselected by detecting $ c_{\rm s} $ signal photocounts can be
reconstructed by the maximum-likelihood method from the conditional
experimental photocount histograms $ f_{\rm ii}(c_{{\rm i}_1},c_{{\rm
i}_2};c_{\rm s}) \equiv f(c_{\rm s},c_{{\rm i}_1},c_{{\rm i}_2}) /
\sum_{c'_{{\rm i}_1},c'_{{\rm i}_2}=0}^{\infty} f(c_{\rm s},c'_{{\rm
i}_1},c'_{{\rm i}_2}) $. We arrive at the following iteration procedure in this
case:
\begin{eqnarray}   
 p^{(j+1)}_{\rm ii}(n_{{\rm i}_1},n_{{\rm i}_2};c_{\rm s})&=& \sum_{c_{{\rm i}_1},c_{{\rm i}_2}=0}^{\infty}
  F^{(j)}_{\rm ii}(c_{{\rm i}_1},c_{{\rm i}_2};c_{\rm s}) T_{{\rm i}_1}(c_{{\rm i}_1},n_{{\rm i}_1}) \nonumber \\
 & & \hspace{-3mm} \times  T_{{\rm i}_2}(c_{{\rm i}_2},n_{{\rm i}_2}),
\label{B2} \\
 F^{(j)}_{\rm ii}(c_{{\rm i}_1},c_{{\rm i}_2};c_{\rm s}) &=& f_{ii}(c_{{\rm i}_1},c_{{\rm i}_2};c_{\rm s}) \Biggl[
  \sum_{n'_{{\rm i}_1},n'_{{\rm i}_2}=0}^{\infty}  T_{{\rm i}_1}(c_{{\rm i}_1},n'_{{\rm i}_1}) \nonumber \\
 & & \hspace{-20mm} \times T_{{\rm i}_2}(c_{{\rm i}_2},n'_{{\rm i}_2}) p^{(j)}_{\rm ii}(n'_{{\rm i}_1},n'_{{\rm i}_2};c_{\rm s}) \Bigr]^{-1},
  \hspace{2mm} j=0,1,\ldots \; . \nonumber
\end{eqnarray}

\section{Identification of the nonclassicality}

For the analyzed postselected 2D idler fields, the NCCa $ C_{K}^{L} $ derived
from the Cauchy--Schwarz inequality and the NCCa $ M_{JKL} $ originating in
non-negative quadratic forms \cite{Agarwal1992} of three variables conveniently
written in the matrix form \cite{Vogel2008,Miranowicz2010,PerinaJr2020a} have
been found the most powerful:
\begin{eqnarray}  
 &C_{K}^{L} = \langle W^L \rangle \langle W^{2K-L}\rangle - \langle W^K\rangle^2 <0,& \nonumber\\
 & K \ge 0, 2K\ge L \ge 0,&
\label{C1} \\
 &M_{JKL} = {\rm det} \langle \left[ \begin{array}{ccc} W^{2J} & W^{J+K} & W^{J+L} \\ W^{K+J} & W^{2K} & W^{K+L} \\
  W^{L+J} & W^{L+K} & W^{2L} \end{array} \right] \rangle < 0, & \nonumber \\
 &J,K,L \ge 0 .&
\label{C2}
\end{eqnarray}
In Eqs.~(\ref{C1}) and (\ref{C2}), we use the notation with vector indices $ K
\equiv (k_{{\rm i}_1},k_{{\rm i}_2}) $ in which $ W^K \equiv W_{{\rm
i}_1}^{k_{{\rm i}_1}} W_{{\rm i}_2}^{k_{{\rm i}_2}} $ and $ K! \equiv k_{{\rm
i}_1}!\, k_{{\rm i}_2}! $.

The NCCa $ C_{K}^{L} $ and $ M_{JKL} $ based on the intensity moments are
translated into the corresponding NCCa $ \bar{C}_{K}^{L} $ and $ \bar{M}_{JKL}
$ written for the probabilities of photon-number (photocount) distributions $
p(k_{{\rm i}_1},k_{{\rm i}_2}) \equiv p(K) $
\cite{Klyshko1996,Waks2004,Waks2006,Wakui2014,PerinaJr2017a} using the mapping
originating in the Mandel detection formula \cite{Perina1991,Mandel1995}:
\begin{equation} 
 \langle W^K\rangle \longleftarrow K!p(K) / p(0,0).
\label{C3}
\end{equation}
We note that the mapping (\ref{C3}) assigns photon numbers and the accompanying
probabilities to the powers of intensity moments. The NCCa for probabilities
indicate not only the global nonclassicality of an analyzed field, they may
also provide the information about the location of the nonclassicality across
the profile of photon-number (photocount) distribution \cite{PerinaJr2020a}.
This can be accomplished by applying the following NCCa $ \bar{C}_{p}(K) $ and
$ \bar{M}_{p}(K) $ that involve the above NCCa $ \bar{C}_{K}^{L} $ and $
\bar{M}_{JKL} $ with the indices obeying specific conditions:
\begin{eqnarray}  
 \bar{C}_p(K) &=& \min_{L,|K-L|\le 1} \{ \bar{C}_{K}^{L} \} ,
\label{C4} \\
 \bar{M}_p(K) &=& \min_{J,L,|J-K|\le 1, |L-K|\le 1} \{ \bar{M}_{JKL} \} ,
\label{C5}
\end{eqnarray}
and $ |K-L|\le 1 $ means that both conditions $ |k_{{\rm i}_j}-l_{{\rm i}_j}
|\le 1 $ for $ j=1,2 $ are fulfilled.

\section{Reconstruction of quasi-distributions of integrated intensities}

An $s$-ordered quasi-distribution $ P_{{\rm ii},s}(W_{{\rm i}_1},W_{{\rm i}_2})
$ of the idler-fields integrated intensities $ W_{{\rm i}_1} $ and $ W_{{\rm
i}_2} $ corresponding to a 2D idler-fields photon-number distribution $ p_{\rm
ii}(n_{{\rm i}_1},n_{{\rm i}_2}) $ is obtained using the following formula
\cite{Perina1991}:
\begin{eqnarray} 
 P_{{\rm ii},s}(W_{{\rm i}_1},W_{{\rm i}_2})&=& \frac{4}{(1-s)^2} \exp\left(-\frac{2(W_{{\rm i}_1}+W_{{\rm i}_2})}{1-s}\right)
 \nonumber \\
 & & \hspace{-12mm} \times \sum_{n_{{\rm i}_1},n_{{\rm i}_2} =0}^{\infty}  \frac{p_{\rm ii}(n_{{\rm i}_1},n_{{\rm i}_2})}{n_{{\rm i}_1}!\, n_{{\rm i}_2}!}
  \left(\frac{s+1}{s-1}\right)^{n_{{\rm i}_1}+n_{{\rm i}_2}}  \nonumber \\
 & & \hspace{-12mm} \times L_{n_{{\rm i}_1}}\left(\frac{4W_{{\rm i}_1}}{1-s^2}\right)
   L_{n_{{\rm i}_2}}\left(\frac{4W_{{\rm i}_2}}{1-s^2}\right) .
\label{D1}
\end{eqnarray}
In Eq.~(\ref{D1}), the symbol $ L_k $ stands for the Laguerre
polynomials \cite{Morse1953}.


%

\end{document}